\documentclass[11pt, letterpaper]{article}

\usepackage{cite}
\usepackage{graphicx}
\usepackage{amsmath}
\usepackage{amssymb}
\usepackage{setspace}
\usepackage{epstopdf}
\usepackage{color}
\usepackage[letterpaper,left=1in,right=1in,top=1in,bottom=1in]{geometry}
\usepackage[linesnumbered,ruled,vlined,longend]{algorithm2e}
\usepackage{multirow}
\usepackage{longtable}
\usepackage{rotating}
\usepackage{threeparttable}
\usepackage{colortbl}
\usepackage{amsthm}
\usepackage[modulo]{lineno}
\usepackage{enumerate}
\usepackage{bm}
\usepackage{algorithmic}

\newtheorem{remark}{Remark}

\theoremstyle{definition}

\title{\Large \bf Distributed economic predictive control of integrated energy systems for enhanced synergy and grid response: A decomposition and cooperation strategy}

\author{\centerline{\normalsize Long Wu$^{a,b}$, Xunyuan Yin$^{c}$, Lei Pan$^{a,\ast}$, Jinfeng Liu$^{b,}$\thanks{Corresponding authors: L. Pan. Email: panlei@seu.edu.cn; J. Liu. Tel: +1-780-492-1317. Fax: +1-780-492-2881. Email: jinfeng@ualberta.ca}}\vspace{5mm}\\
	\centerline{\small $^{a}$ National Engineering Research Center of Power Generation Control and Safety,}\\
	\centerline{\small School of Energy and Environment, Southeast University, Nanjing, 210096, China}\\
	\centerline{\small $^{b}$ Department of Chemical \& Materials Engineering, University of Alberta,}\\
	\centerline{\small Edmonton, Alberta, Canada, T6G 1H9}\\
	\centerline{\small $^{c}$ School of Chemistry, Chemical Engineering and Biotechnology,}\\
	\centerline{\small Nanyang Technological University, 62 Nanyang Drive, Singapore, 637459}}

\allowdisplaybreaks

\begin{document}
	
	\date{}
	
	\maketitle
	\setstretch{1.39}
	
	\begin{abstract}
		
		The close integration of increasing operating units into an integrated energy system (IES) results in complex interconnections between these units. The strong dynamic interactions create barriers to designing a successful distributed coordinated controller to achieve synergy between all the units and unlock the potential for grid response. To address these challenges, we introduce a directed graph representation of IESs using an augmented Jacobian matrix to depict their underlying dynamics topology. By utilizing this representation, a generic subsystem decomposition method is proposed to partition the entire IES vertically based on the dynamic time scale and horizontally based on the closeness of interconnections between the operating units. Exploiting the decomposed subsystems, we develop a cooperative distributed economic model predictive control (DEMPC) with multiple global objectives that regulate the generated power at the grid's requests and satisfy the customers cooling and system economic requirements. In the DEMPC, multiple local decision-making agents cooperate sequentially and iteratively to leverage the potential across all the units for system-wide dynamic synergy. Furthermore, we discuss how subsystem decomposition impacts the design of distributed cooperation schemes for IESs and provide a control-oriented basic guideline on the optimal decomposition of complex energy systems. Extensive simulations demonstrate that the control strategies with different levels of decomposition and collaboration will lead to marked differences in the overall performance of IES. The standard control scheme based on the proposed subsystem configuration outperforms the empirical decomposition-based control benchmark by about 20\%. The DEMPC architecture further improves the overall performance of the IES by about 55\% compared to the benchmark.
		
	\end{abstract}
	
	
	\section{Introduction}
	
	 With the rapid development of advanced hybrid energy grid technologies, the popularity of integration of multiple energy systems has grown to pursue higher energy efficiency and lower environmental costs \cite{ramsebner2021single, arent2021multi}. Integrated energy systems (IESs) with tightly interconnected energy subsystems have emerged as a promising alternative to conventional single input-output energy systems due to their more flexible fashion of energy production and consumption \cite{ramsebner2021single, jing2022integrated}. Typically functioning as a prosumer, an IES comprises customers and various operating units, such as buildings, renewable generation, generators, chillers, heat supply units, energy storage, and other auxiliary units. The close integration of these units through material, information, and energy flows results in a complex process network. While IESs exhibit theoretical energy-efficient and operation-flexible properties, the intensive integration of the units with diversely dynamic characteristics into the unified network brings about complicated issues with dynamics and coordinated control \cite{arent2021multi, wang2018review, sharma2022critical}.
	 
	 As a rising influx of intermittent renewable resources into the power grid, grid response and demand response have attracted considerable attention across residential, commercial, and industrial entities \cite{li2021energy, otashu2020scheduling, fu2022utilizing}. These measures aim to reach electrical supply-demand balance in real time, improving the reliability and flexibility of the grid and the operators' ability to manage risks \cite{groppi2021review}. In particular, the grid-responsive building energy systems as potential participants have been widely investigated. On the one hand, building energy systems can proactively interact with smart grids to facilitate power balance, thanks to their demand flexibility \cite{bay2022distributed}. On the other hand, building energy systems can also provide frequency regulation and demand response services at the request of the grids \cite{wang2019development}. Even systems with small installed capacities can play a prominent role in offering grid/demand response and ancillary services to grid operators through aggregators \cite{obi2020distributed}. It also provides additional profit opportunities for these participants \cite{dowling2017multi}.
	 
	 In this energy context, the assimilation of operating units with diverse functions into a unified system that spans supply and demand sides makes IESs an attractive option for grid response \cite{wang2017review, gao2022multiscale}. The existing work on IESs mainly focuses on scheduling based on steady-state optimization. The approach of multiple time-scale rolling optimizations has been used in scheduling IESs to enhance system robustness and match multi-stage energy markets \cite{li2021multi, fang2022multiple}. In recent work \cite{liu2022bi, qin2022multi}, the dynamic inertial of microturbines was taken into consideration to improve system response. Given that IESs provide multiple energy productions for customers, balancing responses to different energy demands, i.e., integrated demand response, has also been discussed in studies \cite{li2023data, li2021two}.
	 
	 The studies on scheduling in IESs typically assume that the operating units can well reach the prescribed references. However, due to the complex dynamic interactions present in IESs, this places a demanding requirement on the real-time control system \cite{baldea2014integrated, dowling2018economic}. Despite this, research has seldom reported on the detailed dynamics and real-time control of IESs at the second-to-minute time scale. Recently, Wu et al. discussed the full dynamic characteristics of IESs and proposed a multi-time-scale framework based on economic model predictive control to operate a power-cooling IES, where the system was decomposed into three layers based on dynamic response \cite{wu2022economic}. Jin et al. developed a linear distributed tracking model predictive control (MPC) to regulate an off-grid power-heat system with a microturbine and a heat pump by partitioning the system into power and heating subsystems \cite{jin2022power}. Lei et al. separated a combined heat and power system into heat and power divisions and designed a two-layer control scheme with an upper layer for heating control and a lower layer for power control \cite{lei2022research}. Studies by Wu et al. \cite{wu2022economic} and Jin et al. \cite{jin2022power} showed that IESs have multiple time scales of dynamic responses, i.e., the dynamic time-scale multiplicity, which may result in an ill-conditioned control problem when performing optimization-based control schemes \cite{yin2017distributed}. Furthermore, Zhang et al. \cite{zhang2022rapid} and Paiva et al. \cite{paiva2013controllable} examined the dynamic performance of controllable hybrid systems in offering grid response, focusing on systems based on fuel cells and gas turbines, as well as renewable energy, respectively. Notably, similar to these hybrid systems, the operating units in IESs are typically dynamically complementary \cite{wu2022economic}. This characteristic endows IESs with significant potential for precise control over power generation in response to unscheduled power requests from the utility grid. However, since system uncertainty intensifies with the depth of providing response \cite{wang2021disturbance, de2022predictive}, IESs participating in the utility grid response necessitates enhanced capability to coordinate all the units to realize this potential while satisfying the system economical and local customers' energy demands. Due to the complexity of dynamics and structure with numerous variables, a conventional centralized scheme is unsuitable for real-time control of IESs \cite{wu2022economic, jin2022power, christofides2013distributed}. Thus, it is essential to have an effective control strategy that coordinates the dynamic behavior of the units in a distributed fashion. However, there have been fewer efforts made toward distributed coordinated control of complex IESs to unleash their potential for rapid response.
	 
	 Distributed and decentralized control architectures are becoming prevalent in addressing the increasing complexity of structure and a large number of decision variables in emerging energy systems \cite{sharma2022critical, christofides2011networked}. Solar and wind hybrid systems have been extensively studied as practical distributed energy systems. Qi et al. \cite{qi2012distributed} proposed a distributed MPC approach to control solar and wind subsystems, which has been adopted in a similar approach reported in \cite{kong2019hierarchical}. In a hierarchical strategy for a power-heat system, Hu et al. \cite{hu2022novel} designed a distributed MPC for each unit at the bottom layer. Additionally, distributed schemes have been used in heating, ventilation, and air conditioning (HVAC) systems. Rawlings et al. \cite{rawlings2018economic} and Bay et al. \cite{bay2022distributed} developed distributed MPCs for multi-building systems to maintain respective indoor temperatures. Kuboth et al. \cite{kuboth2019economic} proposed a parallel distributed MPC strategy to control heating and water supply subsystems in a hybrid HVAC system. Yang et al. \cite{yang2021model} proposed a two-layer MPC to optimize lighting loads and indoor temperature in a multi-zone building, respectively. Moreover, De Lorenzi et al.\cite{de2022predictive} presented a supervisory MPC for each unit in a power and heating system with multiple buildings for offering demand response. Tang et al. designed a supervisory MPC for chillers and cold storage in an HVAC system for demand response \cite{tang2019model}. A similar supervisor MPC was also reported in \cite{dieulot2015economic} to regulate microturbine, photovoltaics, and battery in a hybrid generation system. One critical point commonly disregarded in many studies on energy systems is that an appropriate subsystem decomposition is a prerequisite and fundamental element of a successful non-centralized optimization framework \cite{yin2019subsystem}. In previous research, systems were usually decomposed based on either the structural boundaries of each operating unit \cite{hu2022novel, de2022predictive, kong2019hierarchical} or the types of energy supplied or consumed by the units \cite{jin2022power, lei2022research}. Apart from these explicit inter-unit connections, systems also have implicit underlying inter- and intra-unit connections resulting from the dynamic variables, including input, disturbance, state, and output. However, most existing studies on the control of energy systems tend to overlook this fact. While such oversights may be acceptable in relatively simple energy systems, they can be problematic for IESs with intricate inter- and intra-unit connections involved in dynamics. Decomposition based on explicit inter-unit connections may lead to improper subsystem configuration, where interconnections between subsystems remain dense, or interactions within subsystems become sparse \cite{yin2022community}. Such insufficient subsystem decomposition can cause non-robustness and deteriorated dynamic performance, as distributed controllers rely on multiple local decision-making agents to coordinate their actions \cite{pourkargar2017impact, daoutidis2019decomposition}. Consequently, achieving synergy between units and the potential for IESs' response as a unified system becomes challenging with non-centralized control frameworks. Despite these issues, a systematic approach to subsystem decomposition in complex energy systems remains relatively unexplored, particularly given the twofold complexity arising from time-scale multiplicity and dynamic interconnectivity.
	 
	 Model predictive control (MPC) has gained favor among extensive control techniques for energy systems due to its prediction and correction mechanisms \cite{rawlings2017model}. In addition to the aforementioned relevant work, MPCs have been designed for optimal operation of multi-zone building aggregators in \cite{khatibi2023towards}, building-storage systems in \cite{touretzky2016hierarchical}, and HVAC systems with concentrated solar power systems in \cite{toub2019model}. Moreover, MPC has been investigated for its applications in district heating/cooling networks and building systems for demand response in \cite{saletti2020development, hu2019price}. It is noteworthy that a new nonlinear MPC with a global economic objective, known as economic MPC (EMPC), has been considered a flexible optimal control tool for smart manufacturing \cite{baldea2014integrated}. Unlike conventional tracking MPC, EMPC provides much flexibility in optimizing the operation of a process for achieving system-wide coordination \cite{ellis2017economic, dowling2018economic}. Preliminary exploration of EMPC in the energy field have been made in the management of power stations \cite{zhang2020zone}, microgrids \cite{alarcon2022economic}, and IESs \cite{wu2022economic}. Despite the success of MPC in managing energy systems, it remains unclear how the decomposed subsystems will impact the design of a distributed architecture when developing a distributed EMPC for IESs that reaches dynamic synergy between all operating units.
	 
	 Based on these observations, it can be found that IESs require an effective real-time control strategy that coordinates all the operating units with diverse dynamics. Nevertheless, there has been limited research on this aspect. Due to the multitude of decision variables, the complexity of system structure, and the multiplicity of time scales, a straightforward application of optimization-based control scheme to IESs will result in a significant computational burden \cite{christofides2013distributed}, reduced robustness \cite{daoutidis2019decomposition}, and an optimization problem that is ill-conditioned \cite{yin2017distributed}. A practical alternative is to partition an IES into several smaller subsystems and then design a distributed framework to control the IES in a modular manner. In this framework, inadequate subsystem configurations can negatively impact control performance in IESs \cite{yin2019subsystem}. Therefore, a generic approach to subsystem decomposition is necessary to effectively handle the time-scale multiplicity and intricate interconnectivity present in the dynamics and structure of IESs. However, IESs currently lack a reliable decomposition method for identifying the optimal subsystem configuration that facilitates the design of distributed cooperation schemes. It is also unresolved how to leverage all the decomposed subsystems to fulfill the operational requirements of electricity, cooling/heating, and profitability in tandem.
	 
	 In order to tackle mentioned issues and achieve dynamically optimal modular management, this paper proposes a vertical-horizontal subsystem decomposition method and cooperative distributed economic MPC (DEMPC) for IESs. We take a grid-connected IES for electricity and cooling supply as the considered system. First, the IES is described as a directed graph that consists of nodes and unidirectional edges to reveal its dynamic variables topology. An adjacency matrix based on Jacobian matrices is then constructed to mathematically represent the directed graph. On this basis, we propose a control-oriented method for decomposing the entire IES into smaller subsystems. This method employs the time-scale separation approach to vertical decomposition of the system and the community detection technique for horizontal decomposition based on interconnections between the operating units. The decomposed subsystems exhibit consistent dynamic responses and strong interactions within each subsystem, but distinct dynamic responses and weak interactions between them. As a result, this approach simultaneously addresses the issues of time-scale multiplicity and structural complexity in IESs. Next, using the decomposed subsystems, we develop a distributed cooperation scheme based on economic MPC to precisely regulate the generated power in response to unscheduled power requests from the grid, while also meeting the local cooling requirements and economic demands of the system. The DEMPC involves multiple sequential and iterative agents with global objectives which cooperate in decision-making by exchanging their latest evaluated information. Consequently, all units in the subsystems are capable of collaborating in real time to attain the specified control objectives. Moreover, the latent influence of the subsystem decomposition on the design of distributed MPC for IESs is also discussed, whereby we provide a basic guideline on subsystem decomposition of complex energy systems. The applicability and effectiveness of the proposed decomposition and cooperation strategy are verified by simulations under varying working conditions.
	 
	 The work presented in this paper has several key contributions as follows:
	 \begin{enumerate}
	 	
	 	\item We introduce a novel directed graph representation of IESs using an augmented Jacobian matrix, which explicitly reveals the underlying interconnectivity between all dynamic variables within the structure of IESs.
	 	
	 	\item We propose a generic method for subsystem decomposition based on the dynamic time scale and closeness of interconnections between the dynamic variables, which simultaneously addresses the complexity of the structure and dynamics of IESs, and efficiently determines an optimal subsystem configuration.
	 	
	 	\item We illustrate how the features of the decomposed subsystems affect the design of distributed MPC for IESs, leading to a control-oriented basic guideline on the decomposition of complex energy systems for modular management.
	 	
	 	\item We develop a scalable distributed cooperation scheme based on economic MPC for IESs, which covers all the decomposed subsystems to achieve system-wide synergy and enhance responsiveness and dynamic performance.
	 	
	 \end{enumerate}
 	In summary, this work contributes novel methods for understanding the structure and dynamics of IESs, decomposing complex energy systems, and designing distributed control schemes for efficient and effective modular management.
 	
 	The organization of the paper is as follows: Section 2 explains the considered IES and control problems; Section 3 accounts for the decomposition approach and relevant discussions; Section 4 develops the DEMPC; Section 5 conducts the simulations; Section 6 brings about conclusions.
 	
	\section{System description and problem formulation}
	
	\subsection{System description}
	
	\begin{figure}[!ht]
		\centering
		\includegraphics[width=0.8\hsize]{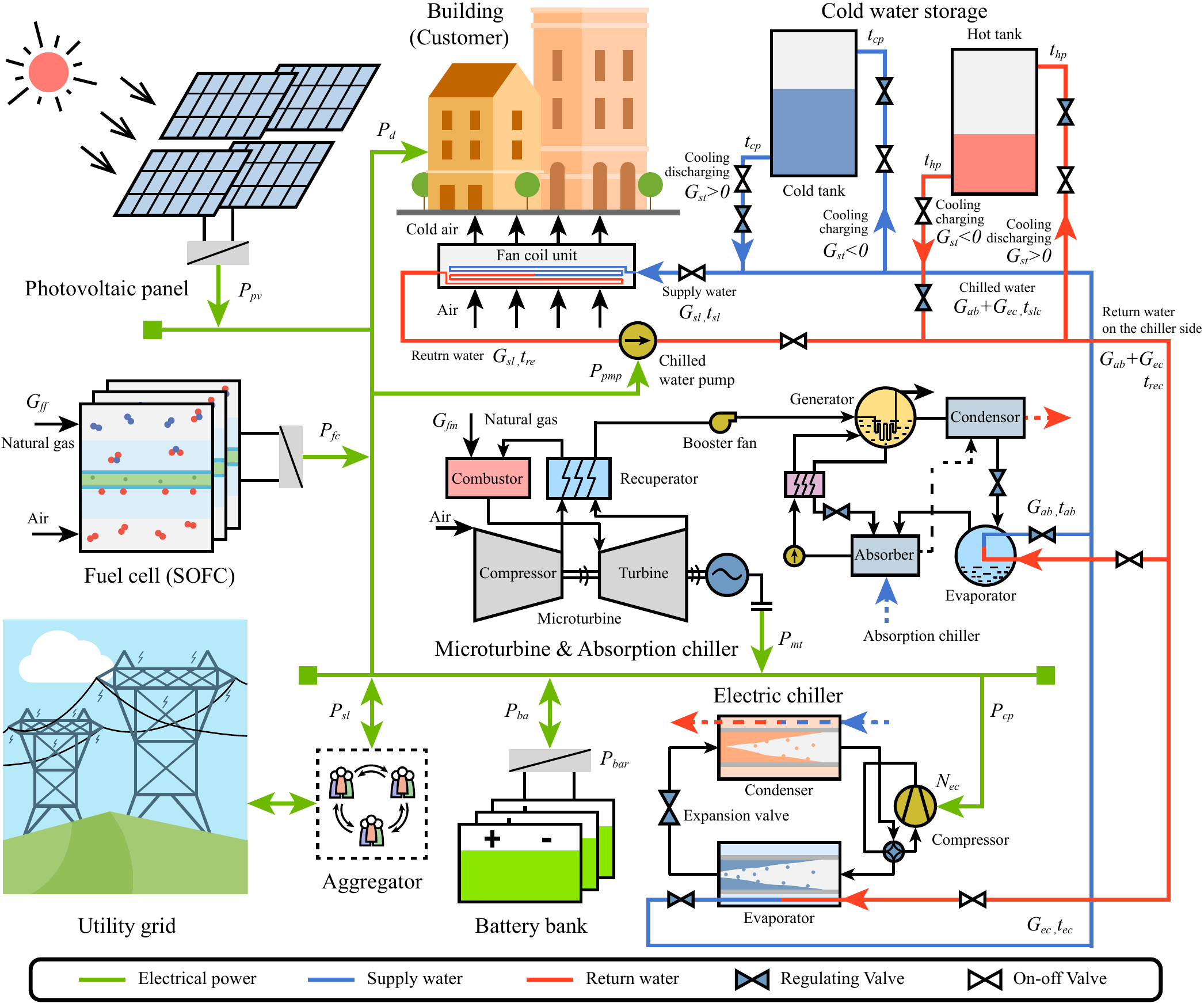}
		\caption{The studied grid-connected integrated energy system.}
		\label{f1}
	\end{figure}
	
	Figure~\ref{f1} shows the grid-connected integrated energy system considered in this work. The IES aims to: (a) satisfy the electric demand in the local microgrid (customer load); (b) maintain the building temperature within the customer-specified range; (c) deliver the extra electricity to the utility grid (hereinafter called the grid), perhaps via an aggregator; (d) increase the IES profitability. The IES is allowed to participate in the grid response. In the case of offering the grid response, the system is required to rapidly regulate its generated power according to the real-time instructions from the grid or aggregator; otherwise, the system will supply electricity to the grid or aggregator according to the prior planned power baseline \cite{dowling2017multi}. Since the instructions are to be an external signal for the IES, we will not distinguish where it comes from after this. Moreover, it is assumed that in this system, when the generated power is insufficient for both the microgrid and grid needs, the electric demand in the microgrid will be met first for the reliability of the local microgrid. Then the surplus/deficient electricity is sent to/from the grid through tie-lines \cite{kaundinya2009grid}. The assumption is made based on the consideration that the grid has other dispatchable generation to keep electrical power flows balanced \cite{gao2022multiscale}.
	
	As shown in Figure~\ref{f1}, the studied IES contains the following operating units: a photovoltaic module (PV), a fuel cell (FC), a battery bank (BA), a microturbine (MT), an absorption chiller (ST), an electric chiller (EC), a cold water storage unit (CS), and a building (customers), which are connected through microgrid or auxiliary pipeline network. On the electricity supply side, the MT and FC utilize natural gas to generate power, while the PV exploits solar energy to produce electrical energy. These units and BA constitute a complementary system to overcome the intermittency issues with renewable resources. Electricity from these units and BA (if in discharge mode) is provided to customers, the water pump, the EC, and BA (if in charge mode) in the microgrid first, and then the grid. On the cooling supply side, the waste flue gas from the combustor of the MT is injected into the AB to produce chilled water. Meanwhile, the EC uses electricity to cool the returned water. Afterward, the chilled water from the chillers is mixed with the cold water from the cold tank (if the CS is in cooling discharge mode) or partly sent to the cold tank (if the CS is in cooling charge mode). The mixture or the rest of the chilled water will be piped to the downstream fan coil unit via the water pump to cool the indoor temperature. The supply water is heated up in the fan coil unit, becoming the return water. If the CS is in cooling discharge mode, the return water is directly piped to the chillers and the hot tank for the next cooling cycle; otherwise, it needs to mix with the water from the hot tank and then go to the chillers. The storage units, the BA and CS, also have to undertake the task of long-term load shifting.
	
	\begin{table} \small		
		\centering
		\caption{Definition of state, input, disturbance, and output variables}
		\label{t0}
		\renewcommand{\arraystretch}{1.1}
		\tabcolsep 5pt
		\begin{tabular}{p{3cm}p{12cm}} \hline
			\multicolumn{2}{l}{\bf{States}} \\
			$I_f$ & Inside current in FC \\
			$G_{H_{2}}$ & Molar flow rate of hydrogen in FC \\
			$p_{O_2}$ & Partial pressure of oxygen in FC \\
			$p_{H_{2}O}$ & Partial pressure of vapor in FC \\
			$p_{H_{2}}$ & Partial pressure of hydrogen in FC \\
			$P_{mtf}$ & Electrical power increment due to fuel flow rate change in MT \\
			$t_{abf}$ & Chilled water temperature increment due to MT exhaust variation in AB \\
			$t_{abw}$ & Chilled water temperature increment due to return flow rate change in AB \\
			$t_{abt}$ & Chilled water temperature increment due to return temperature change in AB \\
			$t_c$ & Condensing temperature in EC \\
			$t_{cs}$ & Condenser shell temperature in EC \\
			$t_{cwm}$ & Mean temperature of cooling water in EC \\
			$t_e$ & Evaporating temperature in EC \\
			$t_{es}$ & Evaporator shell temperature in EC \\
			$t_{ewm}$ & Mean temperature of chilled water in EC \\
			$v_{cap}$ & Capacitor voltage in BA \\
			$C_{soc}$ & Capacity state in BA (state of charge) \\
			$I_{ba}$ & Through current in BA \\
			$C_{sot}$ & Capacity state in CS (state of cold water) \\
			$C_{stc}$ & Heat capacity of water in cold tank in CS \\
			$C_{sth}$ & Heat capacity of water in hot tank in CS \\
			$t_{re}$ & Return water temperature in fan coil unit \\
			$t_{br}$ & Building temperature \\
			\hline
			\multicolumn{2}{l}{\bf{Continuous inputs}} \\
			$G_{ff}$ & Flow rate of feed natural gas in FC \\
			$G_{fm}$ & Flow rate of feed natural gas in MT \\
			$G_{ab}$ & Chilled water flow rate in AB \\
			$N_{ec}$ & Compressor speed in EC \\
			$G_{ec}$ & Chilled water flow rate in EC \\
			$G_{stu}$ & Absolute value of the water flow rate in CS \\
			$P_{bar}$ & Reference signal of output power in BA \\
			\hline
			\multicolumn{2}{l}{\bf{Integer inputs}} \\
			$z_{fc}$ & Switched on/off in FC\\
			$z_{ma}$ & Switched on/off in MT \& AB \\
			$z_{ec}$ & Switched on/off in EC \\
			$z_{st}$ & Cooling charge/discharge in CS \\
			\hline
			\multicolumn{2}{l}{\bf{Disturbances}} \\
			$t_{a}$ & Ambient temperature \\
			$S_{ra}$ & Solar radiation \\
			$P_{d}$ & Electric loads in microgrid\\
			$Q_{o}$ & Other cooling loads \\
			\hline
			\multicolumn{2}{l}{\bf{Outputs}} \\
			$P_{sl}$ & Electricity delivered to the utility grid \\
			$t_{br}$ & Building temperature \\ \hline
		\end{tabular}
	\end{table}
	
	Our previous work \cite{wu2022economic} presented very detailed nonlinear dynamic modeling of each operating unit. Please refer to it for the elaboration and parameters of them. The definition of involved states, manipulated inputs, disturbances, and output variables in the system are listed in Table \ref{t0}. The following part will briefly introduce the system structure.
	
	Specifically, the photovoltaic module with the maximum peak power tracking implementation can be described as a steady-state nonlinear model. The power generated by the photovoltaic panels $P_{pv}$ can be expressed as:
	\begin{equation}
		P_{pv} = n_{pp}n_{sp}\frac{I_{pv,max}V_{pv,max}}{1000}
	\end{equation}
	where $n_{pp}$ and $n_{sp}$ are amounts of parallel and series photovoltaic panels; $I_{pv,max}$ and $V_{pv,max}$ are the current and voltage through each photovoltaic panel under the given disturbances $t_a$ and $S_{ra}$, which satisfies a specific set of nonlinear algebraic equations.
	
	The dynamics of the fuel cell can be represented by five states equations describing the time derivatives of $I_{f}$, $G_{H_2}$, $p_{O_2}$, $p_{H_2O}$, and $p_{H_2}$. Then the electrical power of the fuel cell $P_{fc}$ can be computed as:
	\begin{equation}
		P_{fc} = z_{fc}\frac{V_{fc}I_{fc}}{10^3}
	\end{equation}
	where $V_{fc}$ and $I_{fc}$ denote the output voltage and outside current of the fuel cell, in which $V_{fc} = V_0 - \eta_a - \eta_{c} - \eta_{o}$ where $V_0$, $\eta_a$, $\eta_{c}$, $\eta_{o}$ are the potential voltage and the voltage losses in activation, concentration, and resistance, respectively. All these variables rely on the mentioned five states in the fuel cell, while the state $G_{H_2}$ depends on the input $G_{ff}$.
	
	For the microturbine combined with the absorption chiller, four ordinary differential equations are adopted to characterize its dynamic behavior by states $P_{mtf}$, $t_{abf}$, $t_{abw}$, and $t_{abt}$. $P_{mtf}$ and $t_{abw}$ rely on the the inputs $G_{fm}$ and $G_{ab}$, respectively. Subsequently, the electrical power $P_{mt}$ and supply chilled water temperature $t_{ab}$ produced by the microturbine with the absorption chiller can be evaluated as follows:
	\begin{equation}
		P_{mt} = z_{ma} (P_{mt,0} + P_{mtf})
	\end{equation}
	\begin{equation}
		t_{ab} = t_{ab,0} + t_{abf} + t_{abw} + t_{abt}
	\end{equation}
	where $P_{mt,0}$ and $t_{ab,0}$ are the nominal electrical power and supply chilled water temperature, respectively. The cooling power of the absorption chiller can be calculated by $Q_{ab} = G_{ab}C_{w}(t_{rec} - t_{ab})$, where $C_{w}$ stands for the specific heat capacity of water; $t_{rec}$ is the return water temperature on the chillers side.
	
	Next, a dynamic Thevenin equivalent model of the lithium-ion battery bank is employed in this system. Three ordinary differential equations for three states, $v_{cap}$, $C_{soc}$, and $I_{ba}$, depict its transient performance. Then the power delivered by the battery bank $P_{ba}$ can be obtained from:
	\begin{equation}
		P_{ba} = \frac{V_{ba}I_{ba}}{10^3}
	\end{equation}
	where $V_{ba}$ is the voltage of the battery bank and computed by $V_{ba} = n_{sb}(E_m - v_{cap} - R_{0b}i_{ba})$. In this expression, $n_{sb}$ is the number of series batter cells; $E_m$ accounts for the electrical potential; $R_{0b}$ is the parallel resistance; $i_{ba}$ denotes the current in each parallel battery cell, which is associated with $I_{ba}$ and the input $P_{bar}$. The battery bank is discharged when $P_{ba}>0$, otherwise charged.
	
	The electric chiller involves an evaporator, a compressor, a condenser, and an expansion valve. Its dynamics can be formulated via sixth-order time differential equations composed of six states $t_e$, $t_{es}$, $t_{ewm}$, $t_c$, $t_{cs}$, and $t_{cwm}$. These states relate the two inputs $N_{ec}$ and $G_{ec}$. Next, the consumed electricity of the compressor $P_{cp}$ and chilled water temperature $t_{ec}$ supplied by the electric chiller can be evaluated by:
	\begin{equation}
		P_{cp} = z_{ec}\frac{G_r w_i}{\eta_{cp}}
	\end{equation}
	\begin{equation}
		t_{ec} = 2t_{ewm} - t_{rec}
	\end{equation}
	where $G_r$, $w_i$, and $\eta_{cp}$ are the mass flow rate of refrigerant, the specific power of the compressor, and the compressor efficiency, respectively; $t_{rec}$ is the same return water temperature on the chillers side as the absorption chiller. The evolution of $G_r$, $w_i$, and $\eta_{cp}$ depend on the above six states and two inputs. Similarly, the cooling power of the electric chiller can be attained from $Q_{ec} = G_{ec}C_w(t_{rec} - t_{ec})$.
	
	To proceed, the cold water storage unit consists of a cold and hot tan. Its dynamics can be presented with three states derivative equations of $C_{stc}$, $C_{sth}$, and $C_{sot}$. The cooling power of the storage units can be expressed as $Q_{st} = G_{st}C_w(t_{hp} - t_{cp})$, where $G_{st} = z_{st}G_{stu} + (z_{st} - 1)G_{stu}$ is the water flow rate, $t_{hp} = z_{st}t_{re} + (1 - z_{st})t_{sth}$ denotes the water temperature in the pipe connected to the hot tank, $t_{cp} = z_{st}t_{stc} + (1 - z_{st})t_{slc}$ is the water temperature in the pipe to the cold tank. In these formulations, $z_{st}$ and $G_{stu}$ stand for the integer and continuous inputs, respectively; $t_{re}$ is the return water temperature from the building, which is an state in the fan coil unit; $t_{slc}$ is the chilled water temperature supplied by the chillers; $t_{sth}$ and $t_{stc}$ denote the water temperature in the hot and cold tank, which the three states above can evaluate. When $z_{st} = 1$, the cold water storage is discharged, and $G_{st},Q_{st}>0$, otherwise charged.
	
	In the building and auxiliary pipeline networks, the thermal inertial of the building and fan coil unit are captured by two states with time evolution, $t_{br}$ and $t_{re}$. And the building temperature $t_{br}$, which is one of the system outputs, follows the differential equation below:
	\begin{equation}
		\frac{d t_{br}}{d\tau} = \frac{U_{br}(t_a - t_{br}) - Q_{sl} + Q_{o}}{C_{br}} \label{35}
	\end{equation}
	where $U_{br}$ is the heat transfer coefficient; $t_a$ is the same disturbance as in the photovoltaic module; $Q_{sl}$ accounts for the final cooling power supplied to customers; $Q_{o}$ is another disturbance; $C_{br}$ denotes the building's heat capacity. Further, $Q_{sl}$ can be calculated by $Q_{sl} = G_{sl}C_w(t_{re} - t_{sl})$, where $G_{sl}$ is the mass flow rate of the supply water sent to the building and $G_{sl} = G_{ab} + G_{ec} + G_{st}$, $t_{sl}$ is the final supply water temperature to the building and expressed as:
	\begin{equation}
		t_{sl} = \frac{G_{ab}t_{ab} + G_{ec}t_{ec} + G_{st}t_{cp}}{G_{sl}}
	\end{equation}
	In addition, as illustrated in Figure~\ref{f1}, the energy and mass balance of the supply and return water at different locations in the pipeline are formulated as follows:
	\begin{equation}
		t_{rec} = \frac{G_{sl}t_{re} - G_{st}t_{hp}}{G_{ab} + G_{ec}}
	\end{equation}
	\begin{equation}
		t_{slc} = \frac{G_{ab}t_{ab} + G_{ec}t_{ec}}{G_{ab} + G_{ec}}
	\end{equation}
	\begin{equation}
		G_{all} = G_{ab} + G_{ec} + G_{stu}
	\end{equation}
	where the $t_{rec}$ and $t_{slc}$ denote the aforementioned return water temperature on the chillers side and the supply water temperature from the chillers; $G_{all}$ is the total flow rate of the circulated water in pipeline networks. Then the electric power of the water pump can be obtained from:
	\begin{equation}
		P_{pmp} = \frac{G_{all} g_e H_{pmp}}{10^3\eta_{pmp}}
	\end{equation}
	where $g_e$ is the gravitational acceleration; $H_{pmp}$ and $\eta_{pmp}$ are the hydraulic head and pump efficiency, which is associated with $G_{all}$.
	
	Consequently, taking into account the assumption that the electric loads in the local microgrid $P_d$ will be satisfied first, the electrical power delivered to the utility grid, that is another system output, can be calculated below:
	\begin{equation}
		P_{sl} = P_{pv} + P_{fc} + P_{mt} + P_{ba} - P_{cp} - P_{pmp} - P_{d}
	\end{equation}
	
\begin{figure}[!ht]
		\centering
		\includegraphics[width=0.65\hsize]{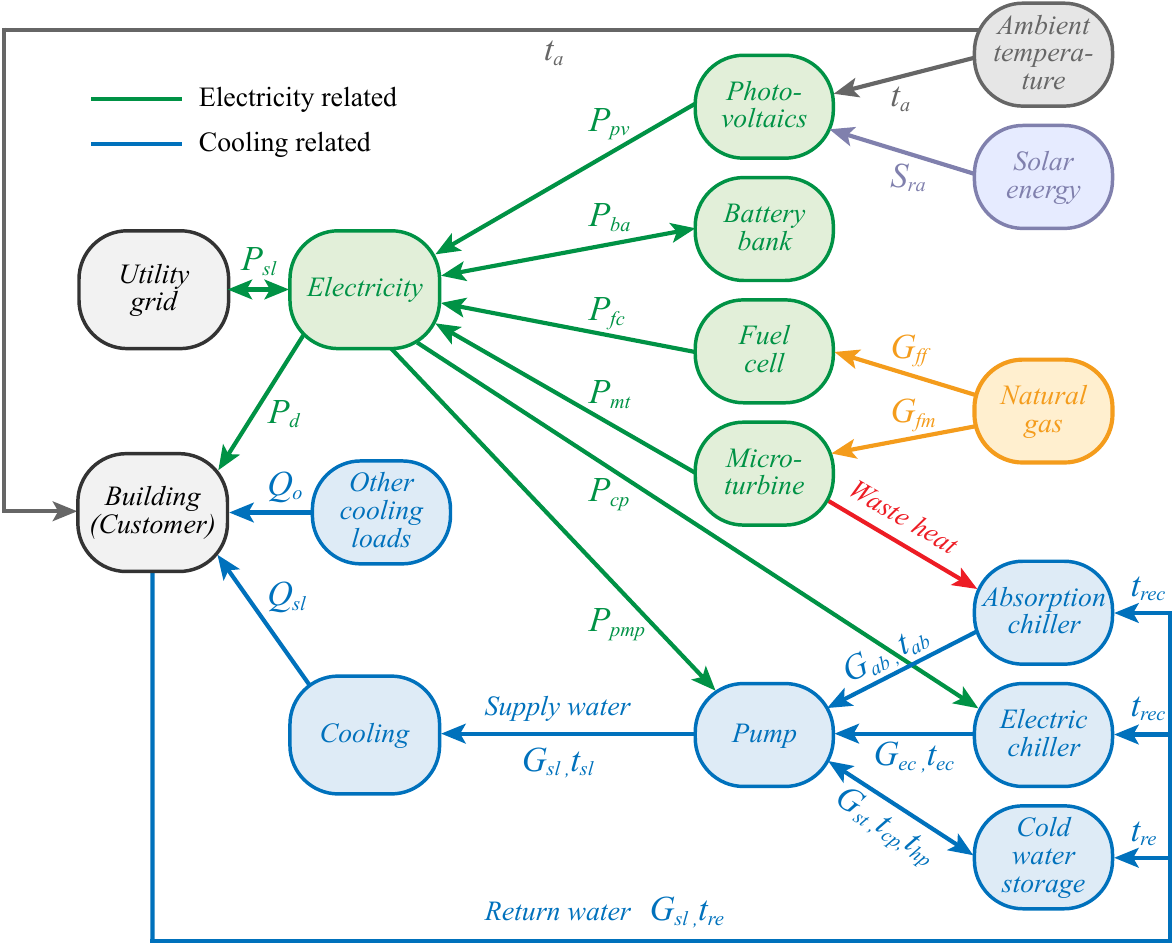}
		\caption{Energy \& material flows and interconnections between operating units within the IES. (Note that the figure illustrates connectivity within the IES at the unit level, without intra-unit interactions.)}
		\label{f2}
	\end{figure}

	To sum up, the dynamic characteristics of the IES are described by 23 nonlinear ordinary differential equations. Figure~\ref{f2} provides an illustration of the interconnections between the operating units within the IES, as well as the energy and material flows that occur throughout the system. Typical variables' values in the IES under nominal conditions are given in Table \ref{t1}.
	
	\begin{table}[!ht] \small
		\centering
		\caption{Typical variables' values under nominal conditions}
		\label{t1}
		\renewcommand{\arraystretch}{1.3}
		\tabcolsep 2pt
		\begin{tabular}{p{2.3cm}p{2.3cm}p{2.3cm}p{2.3cm}p{2.3cm}p{2.3cm}} \hline
			\bf{Variable} & \bf{Value} & \bf{Unit} & \bf{Variable} & \bf{Value} & \bf{Unit} \\ \hline
			$P_{pv}$ & 44 & kW & $Q_{ab}$ & 75 & kW \\
			$P_{fc}$ & 40 & kW & $Q_{ec}$ & 50 & kW \\
			$P_{mt}$ & 80 & kW & $Q_{st}$ & 0 ($-21 \sim 21$) & kW \\
			$P_{ba}$ & 0 ($-40 \sim 40$) & kW & $Q_{sl}$ & 125 & kW \\
			$P_{cp}$ & 12.6 & kW & $t_{ab}$ & 7 & $^{\circ}$C \\
			$P_{pmp}$ & 13.9 & kW & $t_{ec}$ & 7 & $^{\circ}$C \\
			$t_{sl}$ & 7 & $^{\circ}$C & $t_{cp}$ & 7 & $^{\circ}$C \\
			$t_{re}$ & 12 & $^{\circ}$C & $t_{hp}$ & 12 & $^{\circ}$C \\
			\hline
		\end{tabular}
	\end{table}
	
	\subsection{Formulation of control problem}
	
	The grid-connected IES has seven continuous manipulated inputs, four external disturbances (uncontrollable inputs), and two crucial controlled outputs. The system's dynamic behavior is characterized by twenty-three states dispersed in operating units that constitute the IES. Moreover, four integer inputs are involved in the IES to represent the switched on/off of the units and the charge/discharge mode of the cold storage. Then we can define the continuous input vector as $u = [G_{ff}, G_{fm}, G_{ab}, N_{ec}, G_{ec}, G_{stu}, P_{bar}]^T$, the integer input vector as $z = [z_{fc}, z_{ma}, z_{ec}, z_{st}]^T$, the disturbance vector as $\omega = [t_a, S_{ra}, P_d, Q_{o}]^T$, the state vector as $x = [I_f, G_{H_{2}}, p_{O_2}, p_{H_{2}O}, p_{H_{2}}, P_{mtf}, t_{abf}, \\ t_{abw}, t_{abt}, t_c, t_{cs}, t_{cwm}, t_e, t_{es}, t_{ewm}, v_{cap}, C_{soc}, I_{ba}, C_{sot}, C_{stc}, C_{sth}, t_{re}, t_{br}]^T$, and the controlled output vector as $y = [P_{sl}, t_{br}]^T$. Thus the grid-connected IES can be presented by a concise nonlinear state-space model below:
	\begin{subequations} \label{e1}
		\begin{align}
			&\frac{dx}{d\tau}=f(x(\tau),u(\tau), z(\tau), \omega (\tau)) \label{e1a}\\
			&y(\tau)=h(x(\tau),u(\tau), z(\tau), \omega (\tau))
		\end{align}
	\end{subequations}
	where $x\in \mathbb{R}^{n_x}$, $u\in \mathbb{R}^{n_u}$, $z\in \{0,1\}^{n_z}$, $\omega \in \mathbb{R}^{n_d}$, $y\in \mathbb{R}^{n_y}$, and $n_x = 23$, $n_u = 7$, $n_z = 4$, $n_d = 4$, $n_y = 2$ are the number of states, continuous inputs, integer inputs, disturbances, and outputs.
	
	The grid-connected IES can be viewed as a highly complex process network, where the various operating units are tightly interconnected through energy, material, and information flows, as depicted in Figure~\ref{f2}. On the one hand, it can be difficult to manage the intricate interconnectivity between numerous decision variables in IESs with a standard predictive control \cite{christofides2013distributed}. The centralized architecture can cause an undue computational burden \cite{christofides2013distributed}. Additionally, if such optimization-based strategies are used without considering the time-scale multiplicity, the resulting control problem can become ill-conditioned \cite{yin2017distributed}. On the other hand, a non-centralized framework may lead to non-robustness and degraded control performance if the subsystems are not properly configured when dealing with the connectivity in the structure of IESs \cite{daoutidis2019decomposition, yin2019subsystem}. Furthermore, due to the volatile nature of renewable energy and local customers' demands, the IES is susceptible to unstable environmental conditions and electric and cooling demands in the local microgrid. In summary, the primary challenge of real-time coordinated control of IESs is to effectively tackle both the time-scale multiplicity and the strong inter- and intra-unit interactions exhibited in IESs while successfully overcoming external disturbances. The goal is to enable the units to collaborate closely to regulate their generated or consumed electricity in response to the grid's requests while maintaining the building temperature and system profitability. Three evaluation indices of system performance are formulated as follows:
	\begin{subequations} \label{e5} 
		\begin{align}
			J_1 &= \|y_{1}-(1+\xi)y_e^b\|^2 \label{e5a}\\
			J_2 &= \|y_{2}-y_{sp,t}\|^2 \label{e5b}\\
			J_3 &= -\left(p_{mg}\omega_3 + p_{se}y_{1} + p_{cm}\xi_{as} y_e^b - p_f(u_{1}+u_{2}) - p_{pn}\|y_{1} - (1+\xi)y_e^b\|^2\right) \label{e5c}
		\end{align}
	\end{subequations}
	where $\xi$ denotes the regulation capacity factor, which is the ratio of the unscheduled power requested by the grid in real time to the planned baseline power $y_e^b$ previously committed to the grid; $y_{sp,t}$ is the desired building temperature within the acceptable range; $p_{mg}$ and $p_{se}$ are the electricity price in the microgrid and the wholesale electricity price in the grid; $p_{cm}$ is the compensation for offering grid response; $\xi_{as} = |\xi|$ is the absolute value of the regulation capacity factor; $p_f$ is natural gas price; $p_{pn}$ represents fines for failure to offer promised grid response \cite{new2022new}. $J_1$ and $J_2$ reflect deviations of the supplied electricity from the grid's instructions and the building temperature from the customers specified, respectively. $J_3$ evaluates the system's profitability, which equals the negative profit value.
	
	\section{Proposed vertical-horizontal subsystem decomposition}
	
	\begin{figure}[!ht]
		\centering
		\includegraphics[width=0.7\hsize]{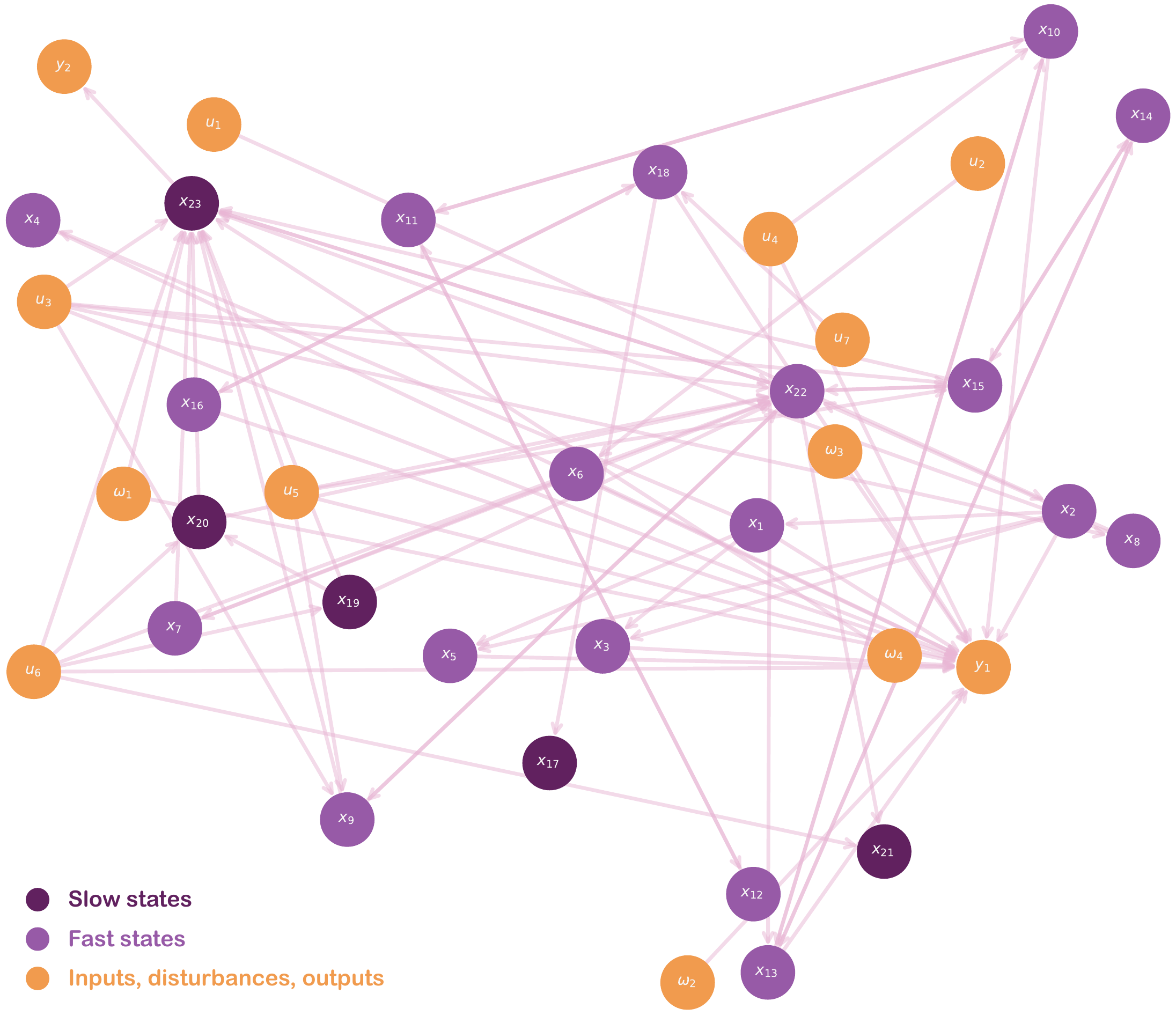}
		\caption{The directed graph of the grid-connected IES: treating variables as nodes and connecting them via directed edges: (1) the dark and light purple nodes account for slow and fast states; (2) the orange nodes stand for manipulated inputs, disturbances (uncontrollable inputs), and controlled outputs; (3) the directed edges represent the existence that the corresponding variable of the start node of the edge directly influences the value or dynamics of the variable the end node indicated.}
		\label{f3}
	\end{figure}
	
	The grid-connected IES can be represented as a directed graph, illustrated in Figure~\ref{f3}. This graph considers the details of inter- and intra-unit connections, including the interactions among input, disturbance, state, and output variables. Compared to Figure~\ref{f2}, Figure~\ref{f3} reveals the underlying network topology of the IES and the interconnections among various variables that the developed control scheme will handle. Distributed and decentralized control approaches are highly effective in coordinating large-scale energy systems with intricate structures and numerous tightly interconnected decision variables \cite{christofides2011networked}. To perform these non-centralized strategies in IESs, determining an optimal subsystem configuration is a crucial first step. Additionally, dynamic time-scale multiplicity must be taken into account when dealing with interconnections between dynamic variables. To this end, this section presents a vertical-horizontal subsystem decomposition framework based on enhanced time-scale separation and community detection techniques to resolve the above difficulties simultaneously.
	
	\subsection{Construction of adjacency matrix for directed graph}
	
	While the directed graph Figure~\ref{f3} presents detailed relationships between input, disturbance, state, and output variables, it can be challenging to use directly to analyze interconnectivity between variables and reach the subsystem decomposition goals. Therefore, in this work, an adjacency matrix is constructed based on Jacobian matrices of IESs to represent the directed graph mathematically.
	
	Before constructing the adjacency matrix, it is important to note the following points in Figure~\ref{f3}. First, since the focus is on interconnections and interactions between different variables (nodes) in the IES, the directed graph does not contain directed self-edges where the start and end nodes are the same. Second, the impact of disturbances on other variables is taken into account, as IESs are vulnerable to uncertain external conditions. Last, according to the definition of directed edges, a directed edge exists between two nodes (i.e., two variables) when the partial derivative of the start variable's function with respect to the end variable is not zero, and vice versa \cite{jogwar2017community}. For instance, there is a directed edge from $u_3$ to $x_{23}$ in Figure~\ref{f3}, which equals to $\partial f_{23}(x,u,z,\omega)/\partial u_3 \neq 0$ where $f_{23}(x,u,z,\omega)$ is the 23rd element of the vector field $f$ in Eq.\eqref{e1}.
	
	Based on these considerations, an augmented Jacobian matrix of the entire IES of Eq.\eqref{e1}, $\tilde{\bf A}_e$, is constructed first as follows:
	\begin{equation}
		{\tilde{\bf A}_e} = \left[ {\begin{array}{ccc}
				\tilde A_{n_x \times n_x} & \tilde B_{n_x \times n_g} & {\bf 0}_{n_x \times n_y}\\
				{\bf 0}_{n_g \times n_x} & {\bf 0}_{n_g\times n_g} & {\bf 0}_{n_g\times n_y}\\
				\tilde C_{n_y \times n_x} & \tilde D_{n_y\times n_g} & {\bf 0}_{n_y\times n_y}
		\end{array}} \right]_{n_v\times n_v}
	\end{equation}
	where $n_v = n_u + n_d + n_x + n_y$ is the total number of input, disturbance, state, and output variables; $n_g = n_u + n_d$ is the number of variables in an augmented input vector $u_g$ consisting of the original input and disturbance vectors, namely $u_g = [u^T, \omega^T]^T$; $\tilde A$, $\tilde B$, $\tilde C$, and $\tilde D$ are following Jacobian matrices of the vector field $f$ and $h$ in Eq.\eqref{e1}:
	\begin{equation}
		\begin{aligned}
			&\tilde A = \frac{\partial f(x,u,z,\omega)}{\partial x}\bigg|_{(x_e,u_e,z_e,\omega_e)}, ~~\tilde B = \frac{\partial f(x,u,z,\omega)}{\partial u_g}\bigg|_{(x_e,u_e,z_e,\omega_e)}\\
			&\tilde C = \frac{\partial h(x,u,z,\omega)}{\partial x}\bigg|_{(x_e,u_e,z_e,\omega_e)},~~\tilde D = \frac{\partial h(x,u,z,\omega)}{\partial u_g}\bigg|_{(x_e,u_e,z_e,\omega_e)}
		\end{aligned}
	\end{equation}
	where $(x_e,u_e,z_e,\omega_e)$ denotes an equilibrium point of the system. $\tilde A$, $\tilde B$, $\tilde C$, and $\tilde D$ reflect whether there exists following directed edges: state-to-state, input/disturbance-to-state, state-to-output, and input/disturbance-to-output, respectively. Suppose the partial derivative indicated by a certain element in these Jacobian matrices is non-zero. Then a directed edge between the corresponding two nodes (variables) should be placed in the directed graph, and vice versa.
	
	\begin{figure}[!ht]
		\centering
		\includegraphics[width=0.55\hsize]{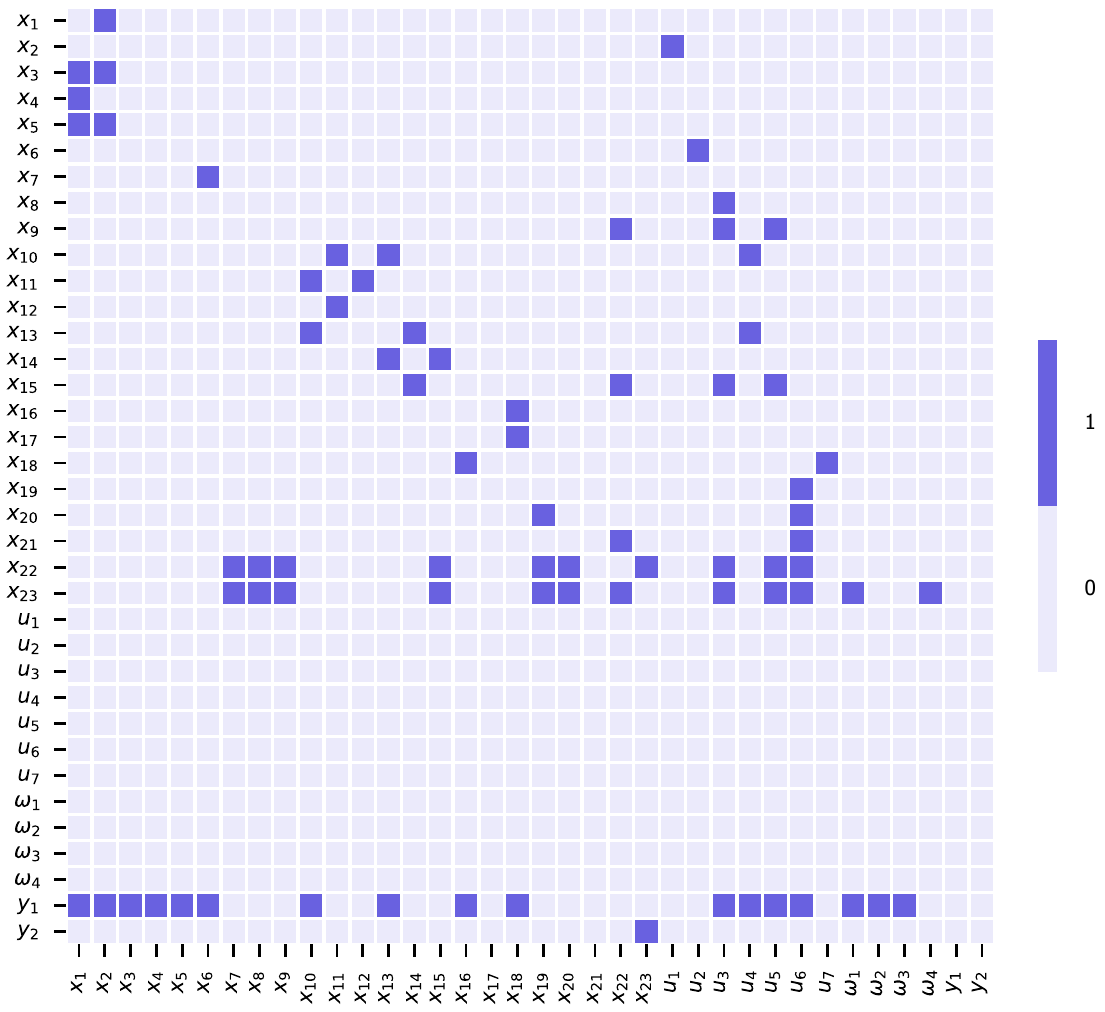}
		\caption{The visualized adjacency matrix of the entire network: a mathematical representation of the directed graph of the IES.}
		\label{f4}
	\end{figure}
	
	Accordingly, the adjacency matrix ${\bf A}_e$ of the entire IES network can be attained by: 
	
	(a) replacing all non-zero elements in $\tilde{\bf A}_e$ with one; 
	
	(b) letting the diagonal elements in $\tilde{\bf A}_e$ be zero since the self-edges are excluded from the graph.
	
	In the adjacency matrix ${\bf A}_e$, therefore, an element located at $i$-th row and $j$-th column is one, denoted as $a^e_{ij} = 1$, when there is a directed edge from node $i$ to $j$, otherwise $a^e_{ij} = 0$. Consequently, the adjacency matrix of the IES under nominal conditions can be calculated and visualized as Figure~\ref{f4}, which will be employed in the vertical-horizontal decomposition of the IES later.
	
	\begin{remark}
		The persistence of external disturbances in IESs poses one of the major challenges to reaching coordination control goals. Thus, the effects of disturbance are incorporated into the directed graph and adjacency matrix. While partitioning disturbance variables may be unnecessary due to their uncontrollability, considering connections between disturbances and systems can help achieve an informative and optimal subsystem decomposition.
	\end{remark}
	
	\subsection{Vertical decomposition based on time-scale separation}
	
	In this section, a time-scale separation approach proposed in our previous work \cite{wu2022economic} is exploited to deal with the dynamic time-scale multiplicity in IESs. The approach is improved by introducing the adjacency matrix, which facilitates the efficient realization of subsystem decomposition. This decomposition leads to the vertical separation of the entire IES into slow and fast subsystems along the time-scale reduction. As a result, these subsystems can align with the system's dynamic response at multiple time scales.
	
	\subsubsection{Singular perturbation formulation of the IES}
	
	\begin{table}[!ht] \small 
		\centering
		\caption{Approximate dominant time constants of states}
		\label{t2}
		\renewcommand{\arraystretch}{1.3}
		\tabcolsep 1pt
		\begin{tabular}{p{3cm}<{\centering}p{1.05cm}<{\centering}p{1.05cm}<{\centering}p{1.05cm}<{\centering}p{1.05cm}<{\centering}p{1.05cm}<{\centering}p{1.05cm}<{\centering}p{1.05cm}<{\centering}p{1.05cm}<{\centering}p{1.05cm}<{\centering}p{1.05cm}<{\centering}p{1.05cm}<{\centering}p{1.05cm}} \hline
			State & $x_1$ & $x_2$ & $x_3$ & $x_4$ & $x_5$ & $x_6$ & $x_7$ & $x_8$ & $x_9$ & $x_{10}$ & $x_{11}$ & $x_{12}$ \\ \hline
			Time constant (s) & 0.8 & 5 & 2.9 & 78 & 26 & 20 & 130 & 80 & 70 & 1.2 & 1.5 & 23.3 \\ \hline
			State & $x_{13}$ & $x_{14}$ & $x_{15}$ & $x_{16}$ & $x_{17}$ & $x_{18}$ & $x_{19}$ & $x_{20}$ & $x_{21}$ & $x_{22}$ & $x_{23}$ & \\ \hline
			Time constant (s) & 1.2 & 1.5 & 19.6 & 6.2 & 14865 & 0.8 & 18000 & 18000 & 18000 & 20 & 12652 & \\ \hline
		\end{tabular}
	\end{table}
	
	To implement the time-scale decomposition, we need to identify the dominant time constant of each state in the IES of Eq.\eqref{e1} to confirm the existence of time-scale multiplicity. Accordingly, the approximate time constants of states that can be viewed as indicators of the time scale of dynamic response are calculated and listed in Table \ref{t2} \cite{wu2022economic}. From Table \ref{t2}, it can be observed that time constants of the states $x_{17}$, $x_{19}$, $x_{20}$, $x_{21}$ and $x_{23}$ are considerably larger than those in the rest of states, which means the dynamic response of the former states are significantly slower than latter ones. Hence, all the states in the IES of Eq.\eqref{e1} can be divided into two categories: slow dynamics $x_s$ and fast dynamics $x_f$. The slow dynamics is composed of the five states, $x_s = [x_{17}, x_{19}, x_{20}, x_{21}, x_{23}]^T$, namely, the capacity states of the battery bank and cold storage $x_{17}$ and $x_{19}$, the heat capacity of contents of cold and hot tanks $x_{20}$ and $x_{21}$, and the building temperature $x_{23}$. If we view the building as a passive energy storage unit, then the slow dynamics purely consist of multiple active and passive energy storage units. On the other hand, the remainder of the IES states constitute the fast dynamics, that is, $x_f = [x_1, x_2, x_3, x_4, x_5, x_6, x_7, x_8, x_9, x_{10}, x_{11}, x_{12}, x_{13}, x_{14}, x_{15}, x_{16}, x_{18}, x_{22}]^T$, which mainly includes diverse generators and chillers such as the fuel cell, microturbine, absorption chiller, electric chiller, and batteries. These operating units typically feature faster dynamic responses than buildings' thermal inertial and capacity states in energy storage units. Afterward, the time constant of $x_5$ and $x_{23}$ are selected as representative values of the time constants corresponding to the fast and slow dynamics, denoted as $\tau^f_{tc}$ and $\tau^s_{tc}$.
	
	Subsequently, the grid-connected IES of Eq.\eqref{e1} can be expressed in terms of singular perturbation formulation \cite{kumar1998singular} by dividing the fast dynamics relevant equations in Eq.\eqref{e1a} by $\tau^s_{tc}$ as follows:
	\begin{subequations} \label{e2}
		\begin{align}
			\frac{d x_s}{d\tau} &= f_s(x_s,x_f,u_s,z,\omega) \\
			\epsilon \frac{d x_f}{d \tau} &= f_f(x_s,x_f,u_s,u_f,z) \label{e2b}\\
			y_s(\tau) &= h_s(x_s) \\
			y_{f}(\tau) &= h_{f}(x_f,u_s,u_f,z,\omega)
		\end{align}
	\end{subequations}
	where $\epsilon$ $(0< \epsilon \ll 1)$ stands for a small dimensionless parameter that is constructed to distinguish distinct time scale in the IES and computed as follows:
	\begin{equation}
		\epsilon =\frac{\tau^f_{tc}}{\tau^s_{tc}} = \frac{26}{12652} = 0.00206 \ll 1
	\end{equation}
	
	Using the adjacency matrix ${\bf A}_e$ that can be viewed as a connectivity map of the system, we can examine connections between the slow/fast dynamics and the continuous inputs. For example, in the adjacency matrix ${\bf A}_e$, the element located at the row of $x_{23}$ and the column of $u_3$ is one, which indicates $u_3$ has a straight impact on $x_{23}$. According to the examination, the continuous inputs in Eq.\eqref{e1} can be separated into two types corresponding to the slow and fast dynamics in Eq.\eqref{e2}: the slow continuous input $u_s = [u_3, u_5, u_6]^T$ that immediately influences the slow dynamics, and the fast continuous input $u_f =[u_1, u_2, u_4, u_7]^T$ that explicitly affects the fast dynamics. Separating the controlled outputs in Eq.\eqref{e1} goes similarly. Consequently, $y_s = y_2$, the building temperature, is the slow output in the system, which depends on the slow dynamics; $y_f = y_1$, the electrical power sent to the grid, is the fast output associated with the fast dynamics. Furthermore, it should be mentioned that, in developing Eq.\eqref{e2}, the proposed adjacency matrix is also used to investigate how the slow and fast states are interconnected and how the inputs affect outputs.
	
	\subsubsection{Establishing slow and fast subsystems based on time-scale separation}
	
	Based on the singular perturbation expression of the IES, we can move forward with time-scale separation for the system. According to the singular perturbation theory \cite{yin2017distributed}, we can obtain the following reduced-order model of the slow subsystem by setting $\epsilon = 0$ in Eq.\eqref{e2}:
	\begin{subequations} \label{e3}
		\begin{align}
			\frac{d x_s}{d\tau} &= f_s(x_s,x_f,u_s,z,\omega) \\
			0 &= f_f(x_s,x_f,u_s,u_f,z) \label{e3b}\\
			y_s(\tau) &= h_s(x_s) \\
			y_f(\tau) &= h_f(x_f,u_s,u_f,z,\omega)
		\end{align}
	\end{subequations}
	where the differential equation of the fast dynamics Eq.\eqref{e2b} becomes an algebraic equation Eq.\eqref{e3b}. This algebraic equation indicates that the slow subsystem does not need to consider the fast dynamics.
	
	To establish the reduced-order fast subsystem, we need to define a stretched time scale, $\tau_f = \tau/\epsilon$, to match the fast dynamics and substitute $\tau_f$ into Eq.\eqref{e2}, then let $\epsilon \rightarrow 0$. The reduced-order model of the fast subsystem, therefore, can be presented below:
	\begin{subequations} \label{e4}
		\begin{align}
			\frac{d x_s}{d \tau_f} &= 0 \label{e4a}\\
			\frac{d x_f}{d \tau_f} &= f_f(x_s,x_f,u_s,u_f,z)\\
			y_{f}(\tau_f) &= h_f(x_f,u_s,u_f,z,\omega)
		\end{align}
	\end{subequations}
	In Eq.\eqref{e4a}, the slow dynamics are treated as constant in the fast subsystem. This implies that during the fast time scale, any minor changes in the slow dynamics are negligible, and information about them can be excluded from the fast subsystem.
	
	The vertical decomposition of the IES has been accomplished. All the variables are divided into the slow and fast subsystems based on their time scale of dynamic response. The dynamics in each subsystem are with a unified time scale, thus addressing the dynamic time-scale multiplicity in the IES.
	
	\begin{remark}
		Formulating the standard singular perturbation expression for IESs is a necessary step in the time-scale separation approach. In this study, we have introduced the use of the adjacency matrix to facilitate this procedure. Resorting to the adjacency matrix, we quickly identify which variables are associated with fast and slow dynamics. This allows us to efficiently determine the arguments on the right-hand side of the singular perturbation expression of Eq.\eqref{e2}. Analyzing system connectivity can be daunting for large-scale energy systems with numerous variables. However, the adjacency matrix makes vertical decomposition readily applicable, simplifying the process.
	\end{remark}
	
	\subsection{Horizontal decomposition based on community detection}
	
	Upon completing the vertical decomposition of the IES, it is observed that the fast subsystem of Eq.\eqref{e4} remains a high-order system comprising eighteen states. In contrast, the slow subsystem of Eq.\eqref{e3} consists of only five states, making it suitable for controller design. This result is reasonable since the IES contains diverse generators, chillers, batteries, and auxiliary units, which exhibit relatively fast dynamic responses compared to a few slow states in the building and energy storage units. Furthermore, due to its rapid dynamic response, the fast subsystem can play a critical role in response to the grid's requests. However, the fast subsystem with tightly interconnected variables remains a complex process network. Hence, this section proposes a subsystem decomposition approach based on community detection for IESs, which is viewed as a horizontal decomposition method aimed at tackling interconnectivity within IES dynamics. By horizontally decomposing the considered network (the fast subsystem), the relevant variables are partitioned further into several smaller subsystems.
	
	\subsubsection{Modularity-based community detection}
	
	Community detection is a powerful technique in graph theory that allows the partition of an entire network into desirable communities, or subsystems, which exhibit strong intra-subsystem interactions but weak inter-subsystem interactions \cite{fortunato2010community}. In this work, we introduce community detection to IESs, enabling us to explore and decompose the underlying interconnections within the system. By employing community detection, we can ensure dense intra-subsystem connections and sparse interconnections between different subsystems, as expected by non-centralized control architectures \cite{yin2022community}. Among the existing community detection tools, the Newman-modularity-based approach has emerged as a dominant method. The core of this approach is an index called modularity, which measures the quality of configured subsystem candidates \cite{leicht2008community}. The typical Newman modularity of a configured subsystem $\Theta$ can be computed by:
	\begin{equation} \label{e6}
		M(\Theta) = \frac{1}{m}\sum_{i=1}^{n_v}\sum_{j=1}^{n_v}\left(a_{ij}-\frac{k^{in}_i k^{out}_j}{m}\right)\delta(c_i,c_j)
	\end{equation}
	where $i,j = 1,\dots,n_v$ are the nodes in the directed graph, $n_v$ is the aforementioned total number of nodes (variables) in the directed graph; $m$ stand for the number of edges; $a_{ij}$ is the $(i,j)$-th element in the adjacency matrix of the considered graph, and as a reminder, $a_{ij} = 1$ if there is a directed edge from $j$-th element to $i$-th element, otherwise $a_{ij} = 0$; $k^{in}_i$ and $k^{out}_j$ denote the number of edges heading for and departing from nodes $i$ and $j$, respectively; $c_i$ and $c_j$ ($c_i, c_j\in \{c_1,\dots,c_{n_v}\}$) are community tags representing the subsystems which the nodes $i$ and $j$ belongs to in the subsystem configuration $\Theta$; $\delta(c_i,c_j)$ is the Kronecker delta function to examine whether the nodes $i$ and $j$ belongs to the same subsystem and calculated as follows:
	\begin{equation} 
		\delta(c_i,c_j) = \left\{ 
		\begin{array}{ccc}
			0, & \mbox{if} & c_i\neq c_j\\
			1, & \mbox{if} & c_i = c_j
		\end{array}\right.
	\end{equation}
	
	In the modularity expression of Eq.\eqref{e6}, the term $(a_{ij}-k^{in}_i k^{out}_j/m)$ indicates the difference between the actual probability of edges between node $i$ and $j$ and the expected probability of edges between them, when all the edges in the network are randomly rewired with $k^{in}_i$ and $k^{out}_j$ fixed. The modularity is the sum of this difference and, therefore, reflects the statistical significance of the subsystem partition $\Theta$. And larger modularity represents better subsystem division, that is, dense connections within subsystems but sparse between different subsystems. Thus, community detection aims to find a subsystem configuration $\Theta$ and the corresponding community tag set of all nodes $\{c_1,\dots,c_{n_v}\}$ that maximizes the modularity in all candidate subsystem partitions, that is:
	\begin{equation} \label{e7}
		M = \max_{c_1,\dots,c_n} M(\Theta)
	\end{equation}
	The following sections will give an effective solution to this optimization problem.
	
	\subsubsection{Realization of community detection for subsystem decomposition}
	
	To perform community detection, we require the adjacency matrix of the target network, i.e., the adjacency matrix of the fast subsystem ${\bf A}_f$. Referring back to the reduced-order model of the fast subsystem described in Eq.\eqref{e4}, the slow dynamics in the fast subsystem are considered as a constant vector but still affect the fast dynamics. This implies that the partial derivative of the slow dynamics with respect to the fast dynamics is zero, while the fast dynamics to the slow dynamics remains unaltered. Taking into account the meaning of the elements in the original adjacency matrix ${\bf A}_e$, we can derive ${\bf A}_f$ by replacing all the elements in the rows related to the slow subsystem in ${\bf A}_e$ with zero, namely letting the elements in the rows of $x_{17}$, $x_{19}$, $x_{20}$, $x_{21}$, $x_{23}$, $u_{3}$, $u_{5}$, $u_{6}$, and $y_2$ to be zero.
	
	Next, we will implement community detection to further decompose the fast subsystem. Mathematically, Eq.\eqref{e7} is an optimization problem computationally difficult \cite{brandes2007modularity}, and so we need an approximation algorithm to solve it. In this study, we employ and improve a heuristic method known as the fast unfolding algorithm \cite{blondel2008fast}, which is acknowledged as an effective and efficient approach to dealing with Eq.\eqref{e7}. The proposed community detection can maximize the modularity $M$ and obtain the configured community tag set $\{c_1,\dots,c_n\}$ nearing global optimum, whereby we can formulate several smaller fast subsystems. The execution of the proposed algorithm is illustrated in Algorithm \ref{a1}.
	\begin{algorithm}[!ht] \small \label{a1}
		\caption{Fast unfolding based community detection for horizontal decomposition}
		
		\textbf{Step 1: Parameter setting}:\\
		Predetermine the upper limit of the number of desired communities (subsystems): $N_c^{u}$.\\
		Specify the terminating condition of the main program loop: $N_l^{l}$, the lower limit of the main program looping $N_l$; $N_m^{l}$, the lower limit of the recurrence $N_m$ of the maximum modularity $M_{max}$.\\
		Create the node vector corresponding to the fast adjacency matrix ${\bf A}_f$: ${\bf v} = [x_1, \dots, x_{n_x}, u_1, \dots, u_{n_u}, \omega_1, \dots, \omega_{n_d}, y_1, \dots, y_{n_y}]$.\\
		Create the community tag vector of ${\bf v}$: ${\bf c} = [c_{x_1}, \dots, c_{x_{n_x}}, c_{u_1}, \dots, c_{u_{n_u}}, c_{\omega_1}, \dots, c_{\omega_{n_d}}, c_{y_1}, \dots,c_{y_{n_y}}]$.\\
		
		Create the initially optimal node and community tag vectors: ${\bf c}_{opt}={\bf c}$ and $ {\bf v}_{opt}={\bf v}$.\\
		Set $N_l=0$, $N_m=0$, $M_{max}=0$.\\
		
		\textbf{Step 2: Main program}:\\
		
		\While{$N_l<N_l^l$ or $N_m < N_m^{l}$}{
			\textbf{Step 2.1: Initialization}:\\
			Randomly reassign the order of nodes in ${\bf v}$, denoted as ${\bf \hat{v}}$; calculate the corresponding adjacency matrix $\hat{\bf A}_f$ via ${\bf A}_f$.\\
			Set $k=0$; let ${\bf c}(k)=[1,2,\dots,n_v]$, i.e., assigning the $i$-th node in $\hat{\bf {v}}$ to the $i$-th community; compute $M(k)$ and the number of community $N_c(k)$.
			
			\textbf{Step 2.2: Modularity maximization}:\\
			\Repeat{$M(k) = M(k-1)$}{
				Set $k = k+1$.\\
				\For{each node $i$}{
					Move $i$ into the neighboring nodes $j$ (including $i$ itself).\\
					Find the maximum $M(k)$; let $i$ stay in the corresponding community.}
				Aggregate the nodes in the same community as a new node.\\
				Update ${\bf c}(k)$, $N_c(k)$, and $M(k)$.}
			
			\textbf{Step 2.3: Community reduction}:\\
			\While{$N_c(k)>N_c^{u}$}{
				Set $k = k+1$.\\
				\For{each node $i$}{
					Move $i$ into the neighboring nodes $j$ (excluding $i$ itself). \\ Find the maximum modularity $M_{ij}(k)$.}
				Find the maximum $M_{ij}(k)$; let $M(k)=M_{ij}(k)$; place $i$ into the corresponding community.\\
				Aggregate the nodes in the same community as a new node.\\
				Update ${\bf c}(k)$, $N_c(k)$, and $M(k)$.}
			
			\textbf{Step 2.4: Community configuration update}:\\
			\eIf{$M(k)>M_{max}$}{
				Update $M_{max}=M(k)$, ${\bf c}_{opt} = {\bf c}(k)$, and ${\bf v}_{opt} = \hat{\bf v}$; reset $N_m$ = 0.}{
				\If{$N_l\geq N_l^l$ and $M(k)=M_{max}$}{
					Update $N_m = N_m+1$.}}
			Update $N_l = N_l+1$.}
		
		\textbf{Step 3: End results}:\\
		Extract the relevant variables in each community (subsystem) from ${\bf c}_{opt}$ and ${\bf v}_{opt}$.
		
	\end{algorithm}
	
	The developed community detection for IESs' horizontal decomposition follows basic ideas of the fast unfolding, i.e., finite moving and aggregation (Step 2.2) \cite{blondel2008fast}. The original fast unfolding is a heuristic algorithm that may find a local optimum. Additionally, the resulting subsystems will be adopted to design a distributed controller. Excessive subsystems would lead to an increase in communication costs between the local controllers. To enhance the fast unfolding method, we have implemented the following modifications:
	
	(a) To reduce the impact of the initial node order on decomposition results, the algorithm now employs a set of random node orders (Step 2.1).
	
	(b) To avoid an excessive number of subsystems, which may increase maintenance costs in the control system, we have defined an upper limit for the number of subsystems (Step 2.3).
	
	(c) To prevent the algorithm from getting stuck in a local optimum, we have set thresholds for the number of main program loops and maximum modularity recurrence (Step 2.4).
	
	Based on these improvements, the proposed community detection can give near global optimal solution, which can efficiently and effectively decompose the target system according to the closeness of interconnections between variables.
	
	\subsubsection{Establishing smaller fast subsystems based on community detection}
	
	\begin{table}[!ht] \small 
		\centering
		\caption{Variables and operating units in the fast subsystem 1, 2, 3.}
		\label{t3}
		\renewcommand{\arraystretch}{1.5}
		\tabcolsep 2pt
		\begin{tabular}{p{2cm}<{\centering}p{4.1cm}<{\centering}p{3.1cm}<{\centering}p{2.6cm}<{\centering}p{4cm}<{\centering}} 
			\hline
			Fast subsystem $i$ & \multirow{2}{4.1cm}[3pt]{\centering States} & \multirow{2}{3cm}[3pt]{\centering Inputs} & \multirow{2}{2.5cm}[3pt]{\centering Outputs} & \multirow{2}{4cm}[3pt]{\centering Involved operating units} \\
			\hline
			\multirow{3}{2cm}[3pt]{\centering 1} & $x_1:I_f$, $x_2:G_{H_{2}}$,$x_3:p_{O_2}$, $x_4:p_{H_{2}O}$, $x_5:p_{H_{2}}$, $x_{16}:v_{cap}$, $x_{18}: I_{ba}$ & \multirow{3}{3cm}[3pt]{\centering $u_1:G_{ff}$, $u_7:P_{bar}$} & \multirow{3}{2.5cm}[3pt]{\centering $y_1:P_{sl}$} & \multirow{3}{4cm}[3pt]{\centering {fuel cell, battery, photovoltaics}} \\
			\multirow{3}{2cm}[3pt]{\centering 2} & $x_6:P_{mtf}$, $x_7:t_{abf}$,$x_8:t_{abw}$, $x_9:t_{abt}$, $x_{15}:t_{ewm}$, $x_{22}:t_{re}$ & \multirow{3}{3cm}[3pt]{\centering $u_2:G_{fm}$} & \multirow{3}{2.5cm}[3pt]{\centering $y_1:P_{sl}$} & \multirow{3}{4cm}[3pt]{\centering {microturbine, absorption \& electric chiller, fan coil unit, photovoltaics}} \\
			\multirow{3}{2cm}[3pt]{\centering 3} & $x_{10}:t_c$, $x_{11}:t_{cs}$, $x_{12}:t_{cwm}$, $x_{13}:t_e$, $x_{14}:t_{es}$ & \multirow{3}{3cm}[3pt]{\centering $u_4: N_{ec}$} & \multirow{3}{2.5cm}[3pt]{\centering $y_1:P_{sl}$} & \multirow{3}{4cm}[3pt]{\centering {electric chiller, photovoltaics}} \\ 
			\hline
		\end{tabular}
	\end{table}
	
	Following the proposed community detection, the fast subsystem of Eq.\eqref{e4} can be divided into three smaller-scale subsystems: the fast subsystem 1, the fast subsystem 2, and the fast subsystem 3, respectively. Table \ref{t3} lists the results of the fast subsystem partition. From Table \ref{t3}, we can observe that the fast subsystem 1 contains the operating units that are only related to electrical power generation, that is, the fuel cell, battery bank, and photovoltaics. The fast subsystem 2 embraces the units relevant to electricity, cooling, fan coil units, and pipeline owing to tight interconnections between the upstream microturbine and the downstream absorption chiller and between the latter and the cooling distribution network. Meanwhile, the chilled water's mean temperature in the electric chiller's evaporator, $x_{15}$, is placed in the fast subsystem 2 since it strongly interacts with supply/return water in the pipeline network. Additionally, only the fast subsystem 2 directly interacts with the slow subsystem's dynamics $x_s$. On the other hand, the internal operational states of the electric chiller are partitioned into the fast subsystem 3. These results demonstrate that the proposed horizontal decomposition can adequately capture the underlying interactions between various variables and then partition them without being constrained by the form of operating units.
	
	Based on the decomposition results and the anticipated distributed controller with global objective functions, we can establish uniform models for the fast subsystems 1, 2, and 3 as follows:
	\begin{equation} \label{e9}
		\begin{aligned}
			\frac{d x_{f_j}}{d \tau_f} &=f_{f_j}(x_{f_j},\bar x_{f_j},u_{f_j},\bar u_{f_j},z)\\
			y_{f_j}(\tau_f) &=h_{f_j}(x_{f_j},\bar x_{f_j},u_{f_j},\bar u_{f_j},z,\omega)
		\end{aligned}
	\end{equation}
	where subscript $f_j$, $j = 1,2,3$, represents the fast subsystem $j$; $x_{f_j}$ is the states vector in the fast subsystem $j$; $\bar x_{f_j}$ denotes a state vector containing other subsystems' states that have an immediate impact on the fast subsystem $j$, or the local decision agent of the fast subsystem $j$ will use; $u_{f_j}$ stands for the manipulated input optimized by the local agent of the fast subsystem $j$; $\bar u_{f_j}$ is an input vector determined by the local controller of other subsystems; $\bar u_{f_j}$ either straightly affects the fast subsystem $j$ or has the information required by the local controller of the fast subsystem $j$ for decision-making. Specifically, 
	
	(a) for the fast subsystem 1, $x_{f_1} = [x_1, x_2, x_3, x_4, x_5, x_{16}, x_{18}]^T$, $\bar x_{f_1} = [x_{f_2}^T, x_{f_3}^T]^T$, $u_{f_1} = [u_1, u_7]^T$, $\bar u_{f_1} = [u_s^T, u_{f_3}^T]^T$, $y_{f_1} = y_1$;
	
	(b) for the fast subsystem 2, $x_{f_2} = [x_6, x_7, x_8, x_9, x_{15}, x_{22}]^T$, $\bar x_{f_2} = [x_s^T, x_{f_1}^T, x_{f_3}^T]^T$, $u_{f_2} = u_2$, $\bar u_{f_2} = [u_s^T, u_{f_3}^T]^T$, $y_{f_2} = y_1$;
	
	(c) for the fast subsystem 3, $x_{f_3} = [x_{10}, x_{11}, x_{12}, x_{13}, x_{14}]^T$, $\bar x_{f_3} = [x_{f_1}^T, x_{f_2}^T]^T$, $u_{f_3} = u_4$, $\bar u_{f_3} = u_s$, $y_{f_3} = y_1$.
	
	It should be mentioned that, first, $\bar x_{f_j}$ and $\bar u_{f_j}$ in the fast subsystem $j$ can be attained via information exchange in the designed control strategy. Second, according to the vertical decomposition results of Eq.\eqref{e4a}, $x_s$ and $u_s$ are still viewed as constants in Eq.\eqref{e9} in the fast time scale. Last, during the horizontal decomposition, we make slight adjustments to the results by sharing $y_1$ and photovoltaics with the fast subsystems 2 and 3. For further details, please refer to Remark \ref{r1}. Finally, the entire IES is decomposed into one slow subsystem Eq.\eqref{e3} and three fast subsystems Eq.\eqref{e9} by the vertical-horizontal decomposition approach.
	
	\begin{remark} \label{r1}
		The proposed decomposition method partitions nodes in energy networks into non-overlapping subsystems, offering managers a clearer view of candidate subsystems. How to use the decomposition results will depend on the problem at hand. In this work, the primary goals of coordinated control of the IES are to accomplish the system-wide synergy and improve the responsiveness, which is significantly correlated with the power delivered to the grid $y_1$. We note that $y_1$ is explicitly associated with the fast subsystem 2, via the generated electricity of the microturbine $P_{mt}$, and the fast subsystem 3, via the power consumption in the electric chiller $P_{cp}$. In addition, $y_1$ is straightly affected by the photovoltaics power $P_{pv}$ that depends on the external disturbances. However, if we directly follow the horizontal decomposition results for the fast subsystem, $y_1$ and the photovoltaics will be placed solely in the fast subsystem 1. This configuration may cause some local control agents with local objective functions, the performance of which is generally slightly inferior to that of the global objective functions \cite{christofides2013distributed}. Therefore, we have shared $y_1$ and photovoltaics with the fast subsystem 2 and 3, as given in Table \ref{t3}, to facilitate developing a distributed cooperation scheme with global objectives. Although we did not entirely follow the initial decomposition results, it provided crucial and adequate guidance for subsystem decomposition.
	\end{remark}
	
	\subsection{Discussions on subsystem decomposition for IESs}
	
	This section will cover how subsystem decomposition affects the design of distributed control architectures in IESs and introduce a basic and generic procedure for decomposing complex energy systems.
	
	\begin{figure}[!ht]
		\centering
		\includegraphics[width=0.7\hsize]{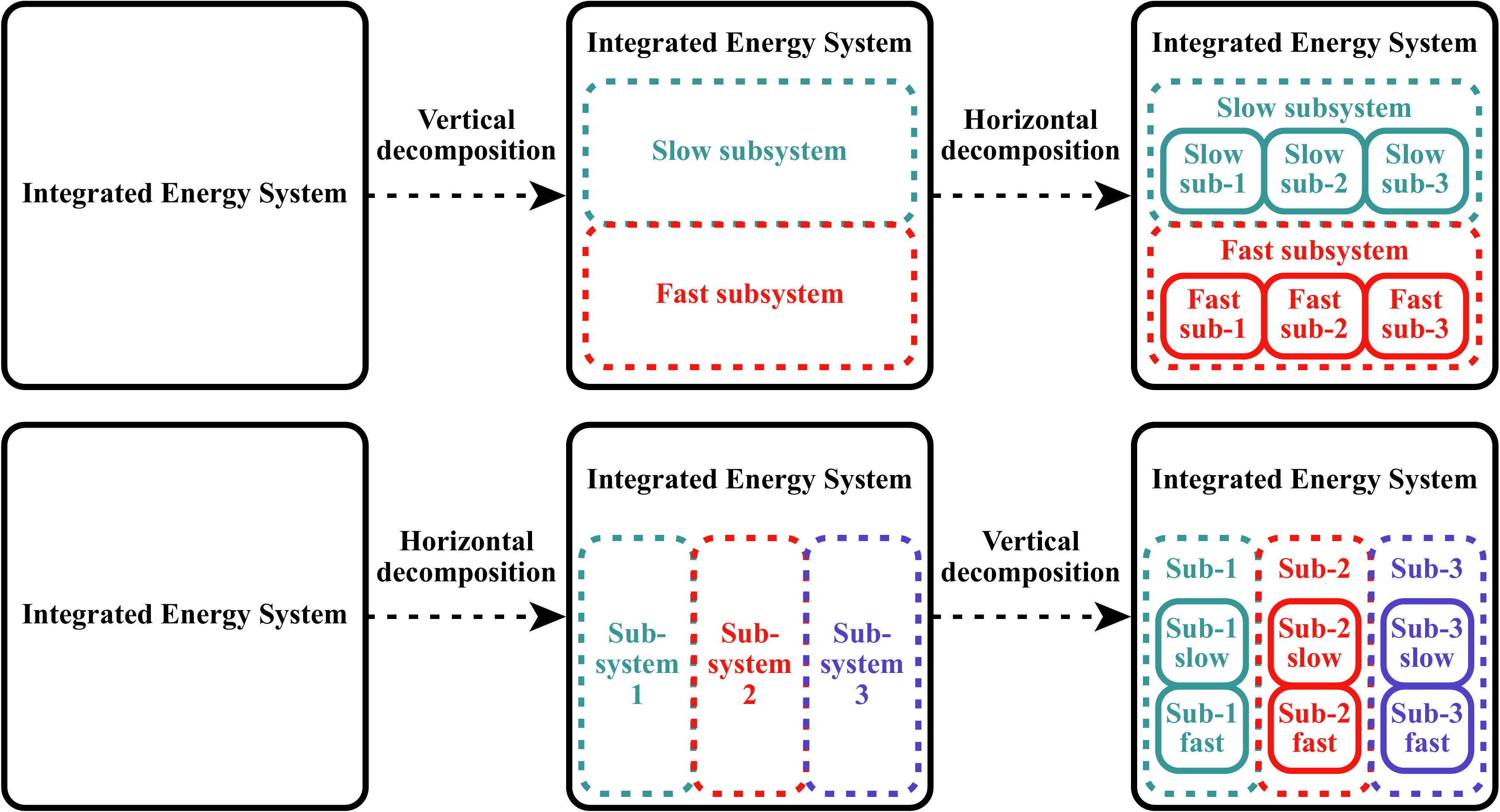}
		\caption{Two possible decomposition procedures: upper: vertical decomposition first and then horizontal decomposition; lower: horizontal decomposition first and then vertical decomposition.}
		\label{f5}
	\end{figure}
	
	In order to handle time-scale multiplicity and structural complexity, the subsystem decomposition method was proposed to partition IESs vertically and horizontally. When vertically partitioning, the information flows sequentially in a top-down direction, typically leading to a sequential distributed control structure \cite{christofides2013distributed}. Horizontal partitioning involves multiple subsystems with mutual information exchange, resulting in an iterative distributed strategy \cite{christofides2013distributed}. In the proposed decomposition framework, it was assumed that the vertical decomposition is performed first and then the horizontal decomposition is implemented in the fast subsystem. Please note that if needed, we can apply the horizontal decomposition to the slow subsystem as well. An alternative to this procedure is to decompose the IES horizontally first and then perform vertical decomposition in each subsystem. The two decomposition procedures can be illustrated in Figure~\ref{f5}. For IESs, the former procedure is recommended. The reasons are twofold: (a) the desired goals of IESs and the level of difficulty of the associated distributed controller design; (b) the size and number of the decomposed subsystems and corresponding communication cost.
	
	\begin{figure}[!ht]
		\centering
		\includegraphics[width=0.7\hsize]{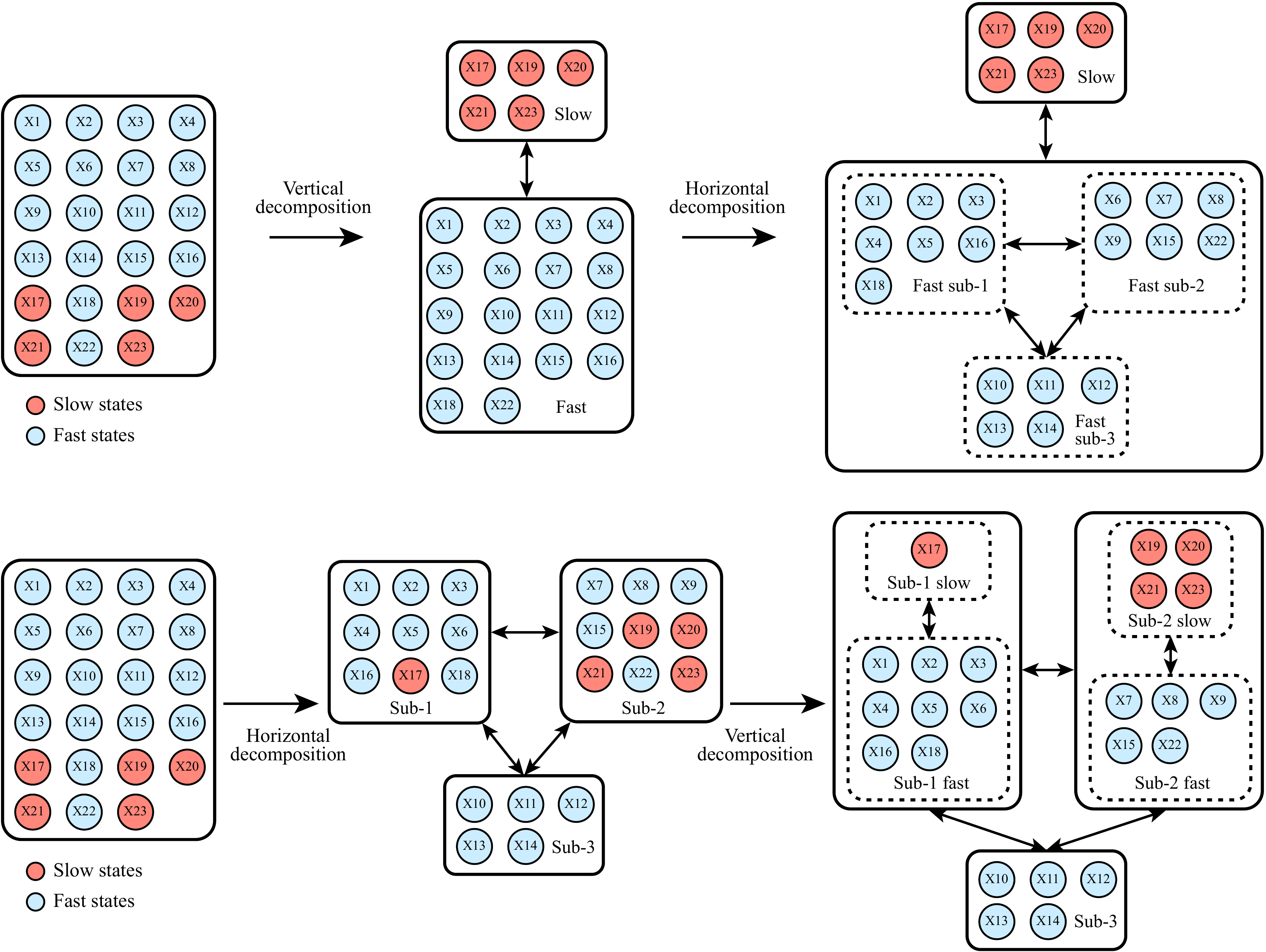}
		\caption{Decomposition results of two decomposition procedures: upper: vertical first and then horizontal; lower: horizontal first and then vertical.}
		\label{f6}
	\end{figure}
	
	Figure~\ref{f6} shows the decomposition results of two procedures applied to the IES. Form Figure~\ref{f6}, it can be seen that performing vertical decomposition and then horizontal decomposition results in four more even subsystems with all the slow states in one subsystem. In this case, the distributed controllers of the slow and fast subsystems simply need to follow the sequential paradigm. An iterative distributed controller can be naturally designed for the fast subsystems such that the operating units can collaborate on the rapid response within a unified time scale. When horizontal decomposition is performed first and then vertical decomposition, it results in five subsystems in which one of them only contains a single slow state, and one does not have any slow states. This decomposition result may be unfavorable from the maintenance and control design perspectives. If we straightly follow this result, non-uniform time scales between and within the subsystems will cause a more complicated iterative distributed controller with asynchronous sampling time \cite{liu2013distributed}. Furthermore, the number of subsystems in this instance is larger than in the former decomposition procedure. Accordingly, the communication cost will be more expensive. An alternative is to artificially merge the two slow subsystems to reduce the number of subsystems. However, it can be a challenging task for a larger energy system with a considerable amount of variables. Moreover, we note that the vertical decomposition first can help create IESs models in different time scales, which favors IESs to take part in energy markets that operate on multiple time scales, see \cite{dowling2017multi}.
	
	\begin{figure}[!ht]
		\centering
		\includegraphics[width=0.8\hsize]{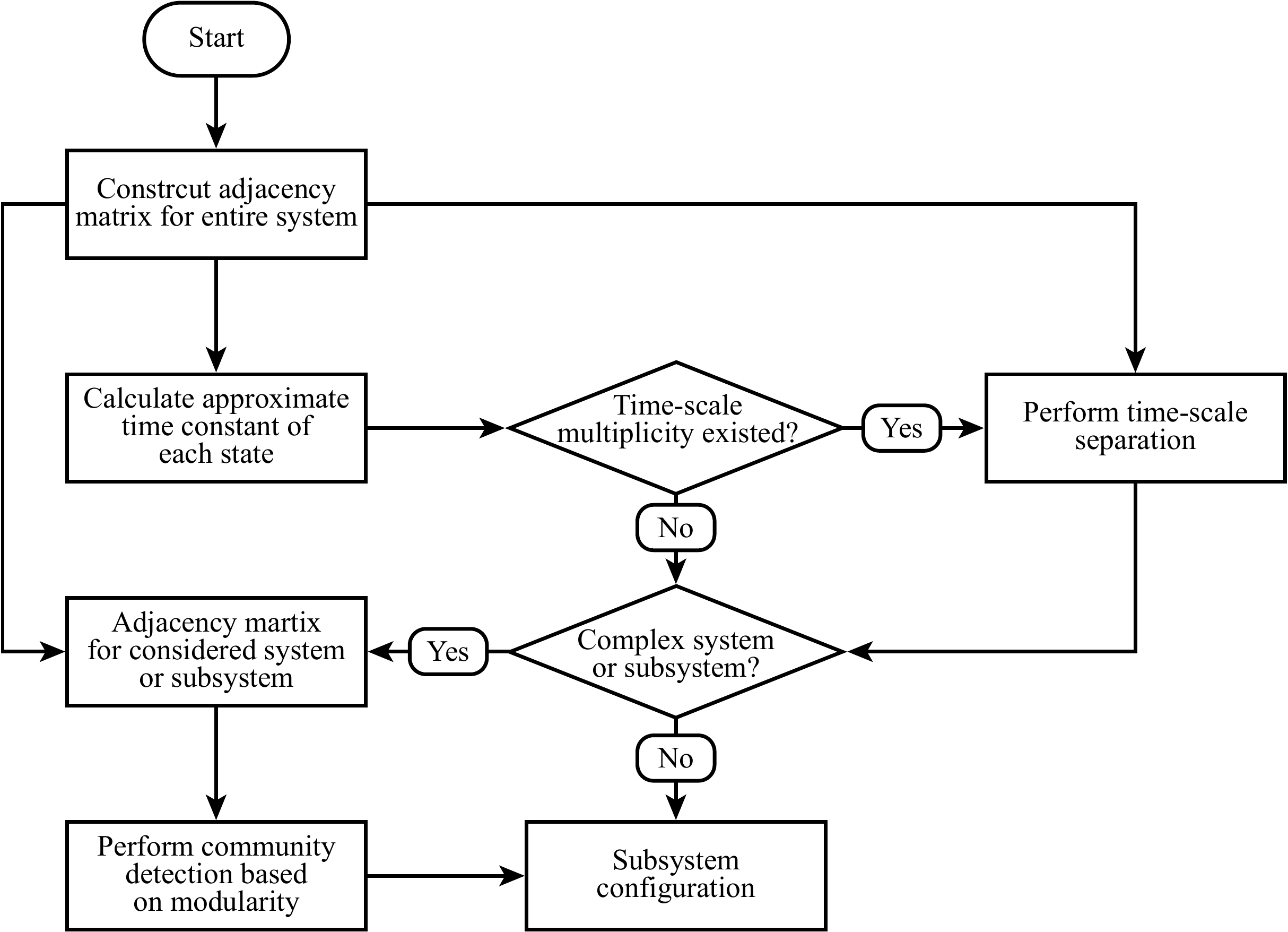}
		\caption{Basic procedure for the generic subsystem decomposition of complex energy systems.}
		\label{f7}
	\end{figure}
	
	Based on these considerations, for IESs or energy systems with time-scale multiplicity and intricate structure, vertical decomposition is a prior task before horizontal decomposition to establish a consistent time scale within a subsystem. Then the horizontal decomposition is implemented to capture the interactions between and within subsystems for further partition. Figure~\ref{f7} illustrates the fundamental steps of the generic vertical-horizontal decomposition approach we propose, which can be applied to other complex energy systems, like virtual power plants and aggregators \cite{ju2022tri, obi2020distributed}. In Figure~\ref{f7}, it should be noted that not all energy systems require both vertical and horizontal decomposition. For instance, large-scale wind and solar farms \cite{kong2019hierarchical} or building groups \cite{bay2022distributed} may not require time-scale separation since they have a unified fast or slow time scale, respectively. How to properly exploit the subsystem decomposition should also consider the optimization problems to be solved. For integrated energy systems across sectors, the decomposition may need to further take into account the potential issues about the attribution of jurisdiction \cite{sneum2021barriers}. The following section will illustrate how to use the features of the decomposed subsystems to design a distributed cooperation scheme based on economic MPC.
	
	\section{Cooperative distributed economic MPC}
	
\begin{figure}[!ht]
		\centering
		\includegraphics[width=0.78\hsize]{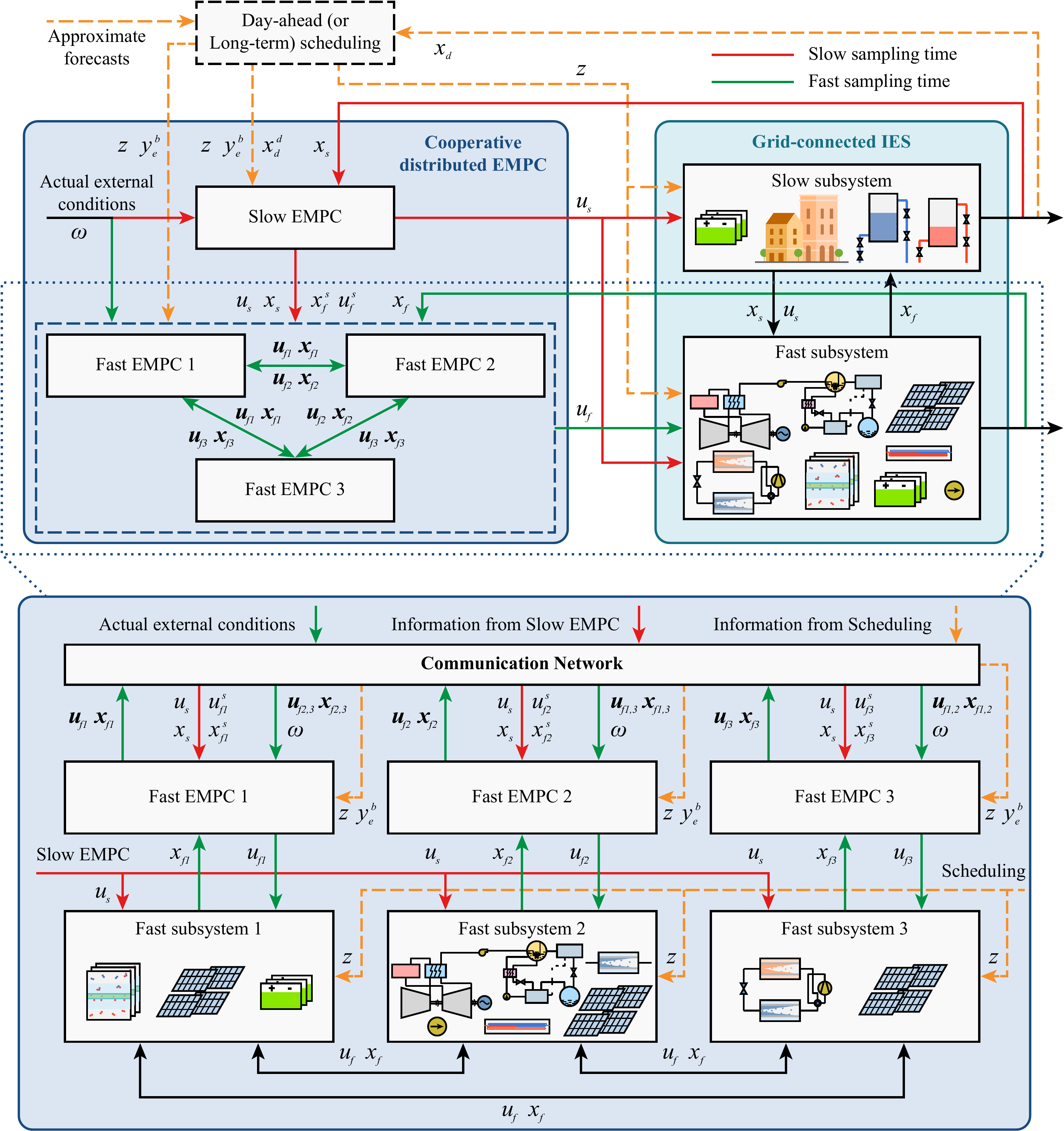}
		\caption{The proposed cooperative distributed economic MPC for the IES.}
		\label{f8}
	\end{figure}

	As previously discussed, a distributed cooperation scheme presents a more favorable alternative to a standard centralized controller in coordinating the operating units within an IES. The effective and reliable management of the IES requires the ability to respond rapidly to changes in the grid's power demands while also satisfying the local electricity and cooling requirements, maximizing profits, and mitigating external disturbances. This section proposes a cooperative distributed economic MPC to address these challenges. Our solution is based on subsystem decomposition and economic MPC, which allows scalable applications in other IESs. The proposed DEMPC framework for the IES is displayed in Figure~\ref{f8}.
	
	The DEMPC also takes into account information exchange with the day-ahead/long-term scheduling (or intraday stages if required). The day-ahead stage will plan the hourly power delivered to the grid and schedule the approximate operational trajectory of the IES based on external conditions forecasts. In this study, we employ a day-ahead optimization adapted from a long-term scheduling approach in our previous work \cite{wu2022economic}. For more information on the scheduling used, please refer to it. The day-ahead scheduling aims to minimize the control objectives described in Eq.\eqref{e5} to determine: (a) the hourly supplied power, i.e., the planned baseline power $y_e^b$ in Eq.\eqref{e5}; (b) the on/off switch of units and charging/discharging of the cold storage, i.e., the integer variable $z$ in Eq.\eqref{e1}; (c) the optimal trajectory references for the capacity states of energy storage for long-term load shifting, i.e., the references for $x_{17}$ ($C_{soc}$) and $x_{19}$ ($C_{sot}$), denoted as a vector $x_d^d$.
	
	Regarding the proposed DEMPC, it consists of four cooperative EMPC based on the vertical-horizontal decomposition results: one slow EMPC for the slow subsystem and three fast EMPCs for the fast subsystems. These local decision-making agents exchange information with each other to coordinate their actions for the dynamic synergy between all the subsystems. In the DEMPC, communication is characterized by one-directional information flow from the slow EMPC to the fast EMPCs at low frequencies, while fast agents exchange information with each other in a high-frequency, mutual manner. The former leads to a sequential distributed EMPC and the latter to an iterative distributed EMPC, which cooperates to leverage the operating units.
	
	Specifically, according to the slow subsystem of Eq.\eqref{e3}, the slow EMPC with about 1-minute sampling time and 10 to 15-minute prediction window is developed to optimize its dynamic behavior under the current external conditions $\omega$ and slow states $x_s$. The optimal actions of the slow inputs $u_s$ and the optimal references for the fast inputs and states, $u_f^s$ and $x_f^s$, can be attained by the slow EMPC that minimizes a global objective function related to $J_1$, $J_2$, and $J_3$. Subsequently, the slow input $u_s$ is immediately applied to the slow subsystem to control it. The needed information, such as $u_s$, $x_s$, $u_f^s$, and $x_f^s$, are sent to the fast EMPCs for further decision-making. During the optimization, the slow EMPC also considers the information about $y_e^b$, $z$, and $x_d^d$ from the day-ahead scheduling.
	
	For the fast EMPCs, the three local control agents with a few seconds sampling time and about 1-minute prediction windows are designed based on the decomposed fast subsystems of Eq.\eqref{e9}. These local agents cooperate to manage the operating units in the fast subsystems under specific external conditions $\omega$ and fast states $x_f$. Fast EMPCs' control actions are coordinated by sharing their latest information and minimizing a joint global objective function that includes $J_1$ and $J_3$. In the communication, the fast EMPCs collect the information from the day-ahead stage for $y_e^b$ and $z$, and from the slow EMPC for $u_s$, $x_s$, $u_f^s$, and $x_f^s$. Additionally, a local fast agent $j$ ($j = 1,2,3$) receives not only other local fast EMPCs' information but also broadcasts information about its newly optimized state and input sequences, ${\bf x}_{f_j}$ and ${\bf u}_{f_j}$, for other local fast agents decision-making. At a sampling time, these three fast EMPCs exchange information and evaluate their actions $u_{f_j}$ iteratively a few times until an iteration limit is approached or the actions of these EMPCs have converged. Consequently, the final optimal fast input $u_{f_j}$ of the fast EMPC $j$ can be obtained by such iterative means, and then $u_{f_j}$ enters the corresponding fast subsystem $j$ to manage the units belongs to it.
	
	\subsection{Sequential distributed slow EMPC}
	
	This section will develop the sequential distributed slow EMPC for the optimal operation of the slow subsystem. At the beginning of the controller design, the control objective of the control scheme needs to be established. For the slow EMPC at a time instance $k$, according to the slow subsystem model of Eq.\eqref{e3} and global operational objectives of Eq.\eqref{e5}, the following control objectives are taken into consideration:
	\begin{subequations}\label{e10}
		\begin{align}
			J_1^s(k) = &\ \alpha_1^s\|y_{f}(k)-(1+\xi(k))y_e^b(k)\|^2 \label{e10a}\\
			J_2^s(k) = &\ \alpha_2^s\|y_{s}(k)-y_{sp,t}(k)\|^2 \label{e10b}\\
			J_3^s(k) = & -\alpha_3^s(p_{mg}(k)\omega_3(k) + p_{se}(k)y_{f}(k) + p_{cm}(k)\xi_{as}(k) y_e^b(k) \notag \\
			&- p_f(k)(u_{f,1}(k)+u_{f,2}(k)) - p_{pn}(k)\|y_{f}(k) - (1+\xi(k))y_e^b(k)\|^2) \label{e10c}\\
			J_4^s(k) = &\ \| x_s^d(k+1) - x_d^d(k+1)\|^2_{R^s} \label{e10d}
		\end{align}
	\end{subequations}
	where $\alpha_w^s$ ($w=1,\dots,3$) and $R^s$ stand for weighting factors and matrix. The objectives of Eqs.\eqref{e10a}-\eqref{e10c} evaluate the systems' performance in tracking the grid's real-time instructions for the supplied electricity, fulfilling the customers' cooling demand, and raising operational revenue, respectively. They are derived from the discretization of the global objectives of Eq.\eqref{e5}. Eq.\eqref{e10d} is a general tracking term for long-term load shifting, in which $x_s^d=[x_{s,1}, x_{s,2}]^T$ is a vector representing the capacity states of battery and cold storage, $x_d^d$ is their optimal references from the day-ahead scheduling.
	
	Taking advantage of these objectives and the slow subsystem model, the sequential distributed slow EMPC is described as follows:
	\begin{subequations} \label{e11}
		\begin{align}
			\min_{u_s, u_f, x_f, y_{sp,t}} & \sum_{i=0}^{N^s_p-1} J_1^s(k+i) + J_2^s(k+i) + J_3^s(k+i) + J_4^s(k+i) \label{e11a} \\ 
			s.t. \ 
			& x_s(k+i+1) = f_s(x_s(k+i),x_f(k+i),u_s(k+i),z(k+i),\omega(k+i)) \label{e11b} \\ 
			& 0 = f_f(x_s(k+i+1),x_f(k+i+1),u_s(k+i),u_f(k+i),z(k+i)) \label{e11c} \\ 
			& y_s(k+i) = h_s(x_s(k+i)) \label{e11d} \\
			& y_f(k+i) = h_f(x_f(k+i),u_s(k+i),u_f(k+i),z(k+i),\omega(k+i)) \label{e11e} \\ 
			& z_r(k+i) u^{min}_r \leq u_r(k+i) \leq z_r(k+i) u^{max}_r \label{e11f} \\
			& \Delta u^{min}_r - V_r \leq \Delta u_r(k+i) \leq \Delta u^{max}_r + V_r \label{e11g} \\
			& x^{min}_r \leq x_{r}(k+i+1) \leq x^{max}_r \label{e11h} \\
			& y_{sp,t}^{min}(k+i) \leq y_{sp,t}(k+i) \leq y_{sp,t}^{max}(k+i) \label{e11i} 
		\end{align}
	\end{subequations}
	where $N^s_p$ accounts for the prediction horizon of the slow EMPC; the subscript $r = s,f$ represents the variables associated with the slow and fast subsystem, and $z_s = {\rm diag} ([z_2,z_3,1])$, $z_f = {\rm diag} ([z_1,z_2,z_3,1])$; the superscripts $min$ and $max$ denote the lower and upper bounds of the relevant variables; $\Delta u_r(k+i) = u_r(k+i) - u_r(k+i-1)$ is the increment of the inputs between two time instants, and $V_r = |z_r(k+i) - z_r(k+i-1)| C_n$ in which $C_n$ is a large enough constant; $V_r$ is used to bypass the constraint of Eq.\eqref{e11g} when $z_r$ changed at time instant $k$; $y_{sp,t}$ is the mentioned desired indoor temperature. Note that $y_e^b$, $x_d^d$, and $z$ are obtained from the day-ahead scheduling.
	
	In the sequential slow EMPC of Eq.\eqref{e11}, the objective function of Eq.\eqref{e11a} is a weighting summation of the above global objectives $J_1^s$, $J_2^s$, $J_3^s$, and $J_4^s$. Eqs.\eqref{e11b}-\eqref{e11e} denote this optimization problem constrained by the discretized slow subsystem model of Eq.\eqref{e3} with slow sampling time $\Delta_s$. Eqs.\eqref{e11f}-\eqref{e11e} are the physical constraints on the inputs and states. Eq.\eqref{e11i} represents the desired indoor temperature set within customer-specified range.
	
	At a sampling time $k$, the slow EMPC will be executed by solving the optimization problem of Eq.\eqref{e11}, whereby the optimal slow input sequence for the entire prediction horizon ${\bf u}_s^*(k)=[u_s^*(k), \dots, u_s^*(k+N^s_p-1)]$ is attained. Meanwhile, the slow EMPC also optimizes the optimal reference sequences for the fast subsystems, ${\bf u}_f^*(k)$ and ${\bf x}_f^*(k)$. And then, the first slow input $u_s^*(k)$ in ${\bf u}_s^*(k)$ is straightly applied to the slow subsystem to regulate its dynamics, while the first elements $u_f^*(k)$ and $x_f^*(k+1)$ in ${\bf u}_f^*(k)$ and ${\bf x}_f^*(k)$ are passed down to the fast EMPCs as references. The remaining elements of ${\bf u}_s^*(k)$, ${\bf u}_f^*(k)$, and ${\bf x}_f^*(k)$ are discarded, which will be re-optimized at next time instant $k+1$ according to the latest external conditions and operational states. Such a procedure is known as the receding horizon implementation \cite{rawlings2017model}. Let us make $u_s(k) := u_s^*(k)$, $u_f^s(k) := u_f^*(k)$, and $x_f^s(k+1) := x_f^*(k+1)$ for the following fast EMPC design.
	
	\subsection{Iterative distributed fast EMPCs}
	
	The proposed method in this section is the three distributed fast EMPCs, which aims to coordinate all the units within the fast subsystems in a synchronized iterative manner to exploit their potential for rapid response. Based on the results of fast subsystem decomposition of Eq.\eqref{e9} and the global control objectives of Eq.\eqref{e5}, unified global objectives for the fast EMPCs at time instant $k$ are formulated as follows:
	\begin{subequations}\label{e12}
		\begin{align}
			J_1^{f_j}(k) = &\ \alpha_1^{f_j}\|y_{f_j}(k)-(1+\xi(k))y_e^b(k)\|^2 \label{e12a} \\
			J_2^{f_j}(k) = & -\alpha_2^{f_j}(p_{mg}(k)\omega_3(k) + p_{se}(k)y_{f_j}(k) + p_{cm}(k)\xi_{as}(k) y_e^b(k) \notag \\
			&- p_f(k)(u_{f_1,1}(k)+u_{f_2,1}(k)) - p_{pn}(k)\|y_{f_j}(k) - (1+\xi(k))y_e^b(k)\|^2) \label{e12b} \\
			J_3^{f_j}(k) = &\ \| v_{f_j}(k) - v_{f_j}^s(k)\|^2_{R_1^{f_j}} \label{e12c} \\
			J_4^{f_j}(k) = &\ \|\epsilon^u_{f_j}(k)\|^2_{R_2^{f_j}} \label{e12d}
		\end{align}
	\end{subequations}
	where superscript and subscript $f_j$ indicate the fast EMPC $j$ or the fast subsystem $j$, and $j = 1,2,3$; $\alpha_w^{f_j}$ and $R_w^{f_j}$ are weighting factors and matrices. Considering that the building temperature is directly influenced by the slow subsystem, the main objectives of the fast EMPCs are to regulate the supplied power in response to the grid's requests and maximize profits. Eqs.\eqref{e12a} and \eqref{e12b} represent these objectives, which also is the discretization of the global objectives of Eqs.\eqref{e5a} and \eqref{e5c}. Eq.\eqref{e12c} is a general tracking objective for ensuring the control system stability. In Eq.\eqref{e12c}, $v_{f_j}(k) = [x_{f_j}(k+1)^T, u_{f_j}(k)^T]^T$ is an augmented vector containing the fast states and inputs, and $v_{f_j}^s(k) = [x_{f_j}^s(k+1)^T, u_{f_j}^s(k)^T]^T$ is a vector consisting of the elements in the optimal references $u_f^s(k)$ and $x_f^s(k+1)$ provided by the slow EMPC. For Eq.\eqref{e12d}, $\epsilon_u^{f_j}(k)$ is a slack variable vector to ensure the stability and iterative convergence of the fast agents.
	
	By taking into account these objectives, the iterative distributed fast EMPC $j$ ($j=1,2,3$) for the fast subsystem $j$ is presented as follows:
	\begin{subequations} \label{e13}
		\begin{align}
			\min_{u_{f_j}, x_{f_j}} & \sum_{i=0}^{N^{f_j}_p-1} J_1^{f_j}(k+i) + J_2^{f_j}(k+i) + J_3^{f_j}(k+i) + J_4^{f_j}(k+i) \label{e13a} \\
			s.t. \ & x_{f_j}(k+i+1) = f_{f_j}(x_{f_j}(k+i),\bar x_{f_j}(k+i),u_{f_j}(k+i),\bar u_{f_j}(k+i),z(k+i)) \label{e13b} \\
			& y_{f_j}(k+i) = h_{f_j}(x_{f_j}(k+i),\bar x_{f_j}(k+i),u_{f_j}(k+i),\bar u_{f_j}(k+i),z(k+i),\omega(k+i)) \label{e13c} \\
			& z_{f_j}(k+i) u^{min}_{f_j} \leq u_{f_j}(k+i) \leq z_{f_j}(k+i) u^{max}_{f_j} \label{e13d} \\
			& \Delta u^{min}_{f_j} - V_{f_j} \leq \Delta u_{f_j}(k+i) \leq \Delta u^{max}_{f_j} + V_{f_j} \label{e13e} \\
			& x^{min}_{f_j} \leq x_{f_j}(k+i+1) \leq x^{max}_{f_j} \label{e13f} \\
			& \Delta u^{min}_{f_j,s} \leq u_{f_j}(k+i) - u_{f_j}^s(k+i) + \epsilon^u_{f_j}(k+i) \leq \Delta u^{max}_{f_j,s} \label{e13g} \\
			& \Delta u^{min}_{f_j,c} \leq u_{f_j}^{(c)}(k+i) - u_{f_j}^{(c-1)}(k+i) \leq \Delta u^{max}_{f_j,c} \label{e13h} \\
			& \Delta u^{min}_{f_j,p} - V_{f_j,p} \leq u_{f_j}(k+i|k) - u_{f_j}(k+i|k-1) \leq \Delta u^{max}_{f_j,p} + V_{f_j,p} \label{e13i} 
		\end{align}
	\end{subequations}
	where subscript and superscript $f_j$ stand for the fast EMPC $j$ for control of the fast subsystem $j$, and $j=1,2,3$; $N^{f_j}_p$ is the prediction horizon of the fast agent $j$; $z_{f_1} = {\rm diag}([z_1,1])$, $z_{f_2} = z_2$, and $z_{f_3} = z_3$ are the integer variable values; the superscripts $min$ and $max$ and the terms $\Delta u_{f_j}$ and $V_{f_j}$ are similar to counterparts in the slow EMPC, which is not repeated; $u^s_{f_j}(k)$ is the input reference for the fast EMPC $j$, which can be extract from $u^s_{f}(k)$ optimized by the slow EMPC; $\epsilon^u_{f_j}(k)$ is the aforementioned slack variable; $u_{f_j}^{(c)}(k)$ represents the evaluated fast input $u_{f_j}$ at iteration $c$ at time instant $k$, in which $c$ indicates $c$-th iteration of the fast EMPCs and $c=1,\dots,c_{max}$, $c_{max}$ is the upper limit of iteration times; $u_{f_j}(k+i|k)$ denotes the predicted fast input $u_{f_j}(k+i)$ at time instant $k$; $V_{f_j,p}$ is used to omit the constraint of Eq.\eqref{e13i} when the current integer variable $z_{f_j}(k)$ is different from the predicted integer variable at last time instant $z_{f_j}(k|k-1)$, and $V_{f_j,p}$ is formulated similar to the aforesaid $V_{f_j}$ and $V_r$. Please note that $y_e^b$ and $z$ are from the day-ahead stage; in light of Eq.\eqref{e4a}, $x_s$ and $u_s$ are given by the slow EMPC and held as constants until the next slow sampling time coming to update them.
	
	In the optimization of Eq.\eqref{e13}, the iterative fast EMPCs are targeted at minimizing the joint global objective function of Eq.\eqref{e13a} that is composed of mentioned $J_1^{f_j}$, $J_2^{f_j}$, $J_3^{f_j}$, and $J_4^{f_j}$. Eqs.\eqref{e13b} and \eqref{e13c} are the constraints of the fast subsystem model, which is attained by discretization of the decomposed fast subsystem of Eq.\eqref{e9} with fast sampling time $\Delta_f$. Eqs.\eqref{e13d}-\eqref{e13f} are the fast system limits on the states and inputs. Eqs.\eqref{e13g}-\eqref{e13f} are the constraints designed to ensure the stability and well iterative convergence of the fast EMPCs. Specifically, Eq.\eqref{e13g} endeavors the fast inputs staying in a neighborhood of the optimized reference by the slow EMPC, which contributes to obtaining an optimal global result and keeping the stability of the slow and fast EMPCs. The slack variable $\epsilon^u_{f_j}$ in Eq.\eqref{e13g} is a penalty term in the objective function. Eq.\eqref{e13h} is adopted to restrict the variation of the fast input between two consecutive iterations at the current sampling time, which helps the fast EMPCs rapidly converge on a joint optimum. Eq.\eqref{e13i} allows the current evaluated fast input $u_{f_j}(k+i|k)$ within a neighborhood of the latest prediction of the current fast input at last time instant $u_{f_j}(k+i|k-1)$. The last element of $u_{f_j}(k+i|k-1)$, i.e., $u_{f_j}(k+N^{f_j}_p-1|k-1)$, is one step zero-order extrapolation from its penultimate elements. Eq.\eqref{e13i} is also utilized to promote fast convergence of the distributed agents towards a joint optimum during iterations. $\Delta u^{min}_{f_j,s}$, $\Delta u^{max}_{f_j,s}$, $\Delta u^{min}_{f_j,c}$, $\Delta u^{max}_{f_j,c}$, $\Delta u^{min}_{f_j,p}$, and $\Delta u^{max}_{f_j,p}$ are tuning parameters in Eqs.\eqref{e13g}-\eqref{e13i}. If they are set relatively small, it helps Eq.\eqref{e13} to converge fast but may bring about a slightly conservative optimization result; conversely, if they are set relatively large, it may expand the feasible region of Eq.\eqref{e13} but need a longer convergent time.
	
	At a sampling time $k$, three fast EMPCs of Eq.\eqref{e13} will be performed iteratively to evaluate their control actions and corresponding states while distributing their latest optimization results to each other for cooperative decision-making. At $c$-th iteration ($1<c<c_{max}$): 
	
	(a) the fast EMPC 1, 2, and 3 exchange their newly optimized input and state sequences at the last iteration $c-1$, i.e., ${\bf u}_{f_j}^{(c-1)}(k)=[u_{f_j}^{(c-1)}(k), \dots, u_{f_j}^{(c-1)}(k+N^{f_j}_p-1)]^T$ and ${\bf x}_{f_j}^{(c-1)}(k)=[x_{f_j}^{(c-1)}(k+1), \dots, x_{f_j}^{(c-1)}(k+N^{f_j}_p)]^T$;
	
	(b) based on the latest ${\bf u}_{f_j}^{(c-1)}(k)$ and ${\bf x}_{f_j}^{(c-1)}(k)$, the fast EMPCs' optimization problems of Eq.\eqref{e13} are solved in a parallel manner to evaluate their input and state sequences ${\bf u}_{f_j}^{(c)}(k)$ and ${\bf x}_{f_j}^{(c)}(k)$ at the current iteration $c$;
	
	(c) and then go to the next iteration $c+1$; again, the fast EMPC 1, 2, and 3 will share their information about input and state sequences ${\bf u}_{f_j}^{(c)}(k)$ and ${\bf x}_{f_j}^{(c)}(k)$ with each other for preparing the iteration;
	
	(d) such iteration procedure will be repeated until either reaching the iteration upper limit $c_{max}$ or all the fast EMPCs converged, i.e., $(J_{f_i}^{(c)}(k) - J_{f_i}^{(c-1)}(k))/J_{f_i}^{(c-1)}(k) \leq \psi$ wherein $\psi$ is a tunable convergent threshold, and $J_{f_i}^{(c)}(k) = \sum_{i=0}^{N^{f_j}_p-1} J_1^{f_j,(c)}(k+i) + J_2^{f_j,(c)}(k+i) + J_3^{f_j,(c)}(k+i) + J_4^{f_j,(c)}(k+i)$, that is, the calculated cost function of Eq.\eqref{e13a} at $c$-th iteration.
	
	Once the last iteration is completed, the fast EMPCs will obtain the finally optimal fast input sequences at the time instant $k$, denoted as ${\bf u}_{f_j}^*(k)=[u_{f_j}(k), \dots, u_{f_j}(k+N^{f_j}_p-1)]^T$. Subsequently, the first element $u_{f_j}(k)$ in ${\bf u}_{f_j}^*(k)$ will be applied to the fast subsystem $j$ to manage it. The rest of the elements in the optimized sequences ${\bf u}_{f_j}^*(k)$ are retained and extrapolated one step as the predicted input $u_{f_j}(k+i+1|k)$ in Eq.\eqref{e13i} for the next time instant optimization. When the next time instant $k+1$ comes, the above iterative receding horizon implementation will be repeated according to the newest external conditions and fast subsystem states.
	
	\begin{remark}
		It should be mentioned that the microturbine combined with the absorption chiller in the fast subsystem 2 generally has a slightly slower dynamic response than the operating units in the other fast subsystems. Thus, the fast EMPC 1, 2, and 3 may have different prediction horizons $N^{f_j}_p$ ($j=1,2,3$) in Eq.\eqref{e13} to cover their major dynamics. In this case, for the fast EMPC $l$ ($l=1,2,3$), cut-off and zero-order extrapolation are used to make the received ${\bf u}_{f_j}^{(c)}(k)$ and ${\bf x}_{f_j}^{(c)}(k)$ ($j \neq l$) with the same length as $N^{f_l}_p$.
	\end{remark}
	
	\section{Simulation and comparison}
	
	This section will apply the proposed DEMPC to the grid-connected IES. We compare its performance with several control schemes using simulations on a machine with a 2.60 GHz Intel Core i7-10750H processor and 16 GB RAM. All the optimization problems are solved by IPOPT (3.12.3) and BONMIN (1.8.4) solvers, which are implemented using the CasADi platform (3.5.5) \cite{andersson2019casadi} in Python (3.7.6).
	
	\subsection{Supervisory MPC architecture for comparison}
	
	\begin{figure}[!ht]
		\centering
		\includegraphics[width=0.6\hsize]{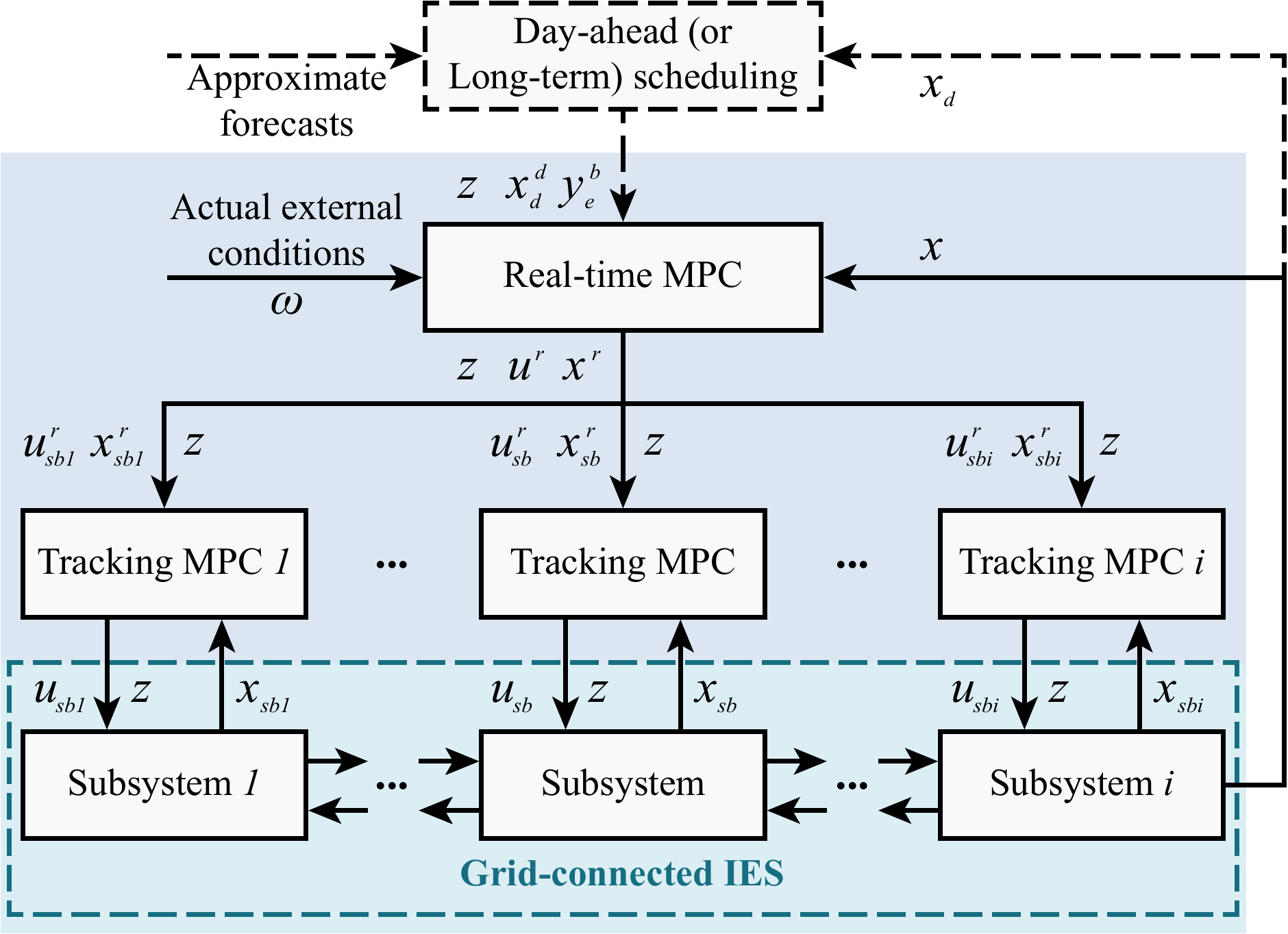}
		\caption{The supervisory MPC architecture for comparison.}
		\label{f9}
	\end{figure}
	
	Since a centralized MPC is inadequate for complex IESs, this study adopts a supervisory MPC architecture for comparative analysis, as illustrated in Figure~\ref{f9}. This architecture has been extensively employed in energy systems to optimize their operations \cite{dieulot2015economic, tang2019model, de2022predictive, qi2011supervisory}. In the supervisory MPC, the entire system is also partitioned into a few subsystems. The day-ahead scheduling is the same as used in the DEMPC. The high-level real-time MPC coordinates the entire system by minimizing a global objective based on Eq.\eqref{e5}. According to the given external conditions $\omega$ and system states $x$, the real-time MPC will evaluate the optimal trajectories of the entire system's inputs and states, $u^r$ and $x^r$. And then dispatch them, as the prescribed references, to the local tracking MPCs that do not communicate with each other. The low-level decentralized tracking MPCs are viewed as local regulators of the respective subsystems. With the new subsystem feedback $x_{sb}$, the local MPCs will track their respective references $u_{sb}^r$ and $x_{sb}^r$ by minimizing local tracking objectives, which decides the final inputs $u_{sb}$ that enters the subsystems. For comparison, we introduce three different subsystem configurations to the supervisory MPC as follows:
	
	(a) the subsystem is divided by straightly following the results of the proposed vertical-horizontal subsystem decomposition;
	
	(b) the subsystem is partitioned in terms of the operating unit: the fuel cell, microturbine with the absorption chiller, electric chiller, cold storage unit, and battery bank are treated as individual subsystems, respectively;
	
	(c) the subsystem is decomposed by considering whether the units are to generate electricity or supply cooling: one subsystem is composed of the fuel cell, microturbine, and battery bank; another subsystem consists of the absorption chiller, electric chiller, and cold storage unit.
	
	For convenience, let us refer to the proposed DEMPC as Problem 1 (or P1), the supervisory MPC based on the subsystem configuration (a) as Problem 2 (or P2), the supervisory MPC based on (b) as Problem 3 (or P3), the supervisory MPC based on (c) as Problem 4 (or P4).
	
	\subsection{Parameter and scenario settings}
	
	For the proposed DEMPC, the sampling time and prediction horizon are listed in Table \ref{t4}. In particular, the day-ahead stage's sampling time and optimization horizon are set as 1 hour and 24 hours. The limit on iterations $c_{max}$ and convergent threshold $\psi$ in the DEMPC are chosen as $c_{max}=12$ and $\psi=5\%$. For the compared supervisory MPCs, the sampling time of the real-time MPC is 60s, and the local tracking MPCs are 5s, which is equal to the DEMPC. The prediction horizon of the real-time MPC is also 12. The local tracking MPCs' prediction horizons are either 10 or 12, which basically follow the fast EMPCs' choice of them. 
	
	\begin{table}[!ht] 
		\centering
		\caption{Sampling time and prediction horizon of the DEMPC}
		\label{t4}
		\renewcommand{\arraystretch}{1.3}
		\tabcolsep 2pt
		\begin{tabular}{p{3.5cm}<{\centering}p{2cm}<{\centering}p{2cm}<{\centering}p{2cm}<{\centering}p{2cm}<{\centering}} \hline
			EMPCs & Slow & Fast 1 & Fast 2 & Fast 3 \\ \hline
			Sampling time (s) & 60 & 5 & 5 & 5 \\ \hline
			Prediction horizon & 12 & 10 & 12 & 10 \\ \hline
		\end{tabular}
	\end{table}
	
	\begin{figure}[!ht]
		\centering
		\includegraphics[width=0.8\hsize]{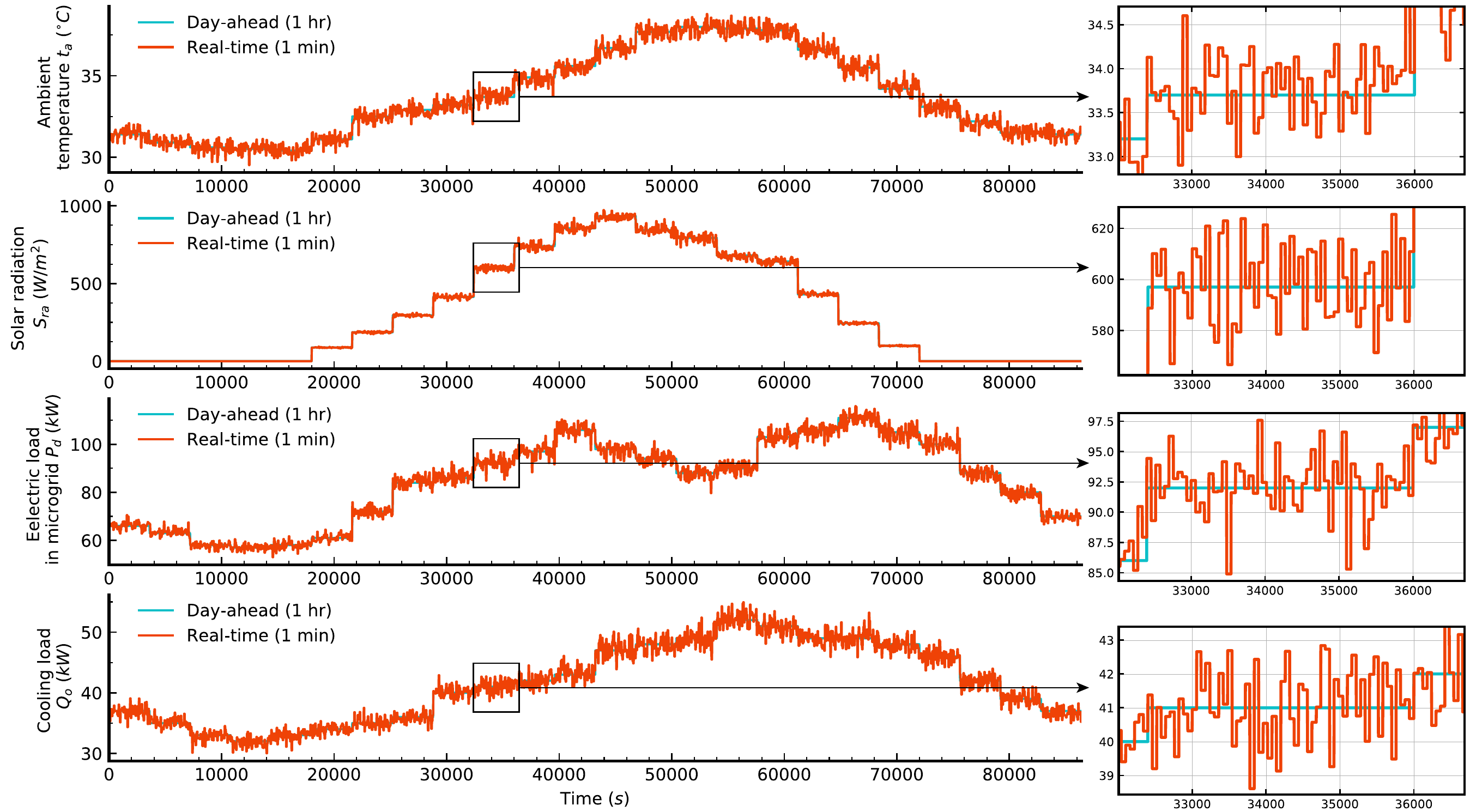}
		\caption{Time evolution of external conditions: the day-ahead predicted curves used in the day-ahead scheduling; the actual real-time curves used in the real-time controls, i.e., Problems 1-4.}
		\label{f10}
	\end{figure}
	
	\begin{figure}[!ht]
		\centering
		\includegraphics[width=0.7\hsize]{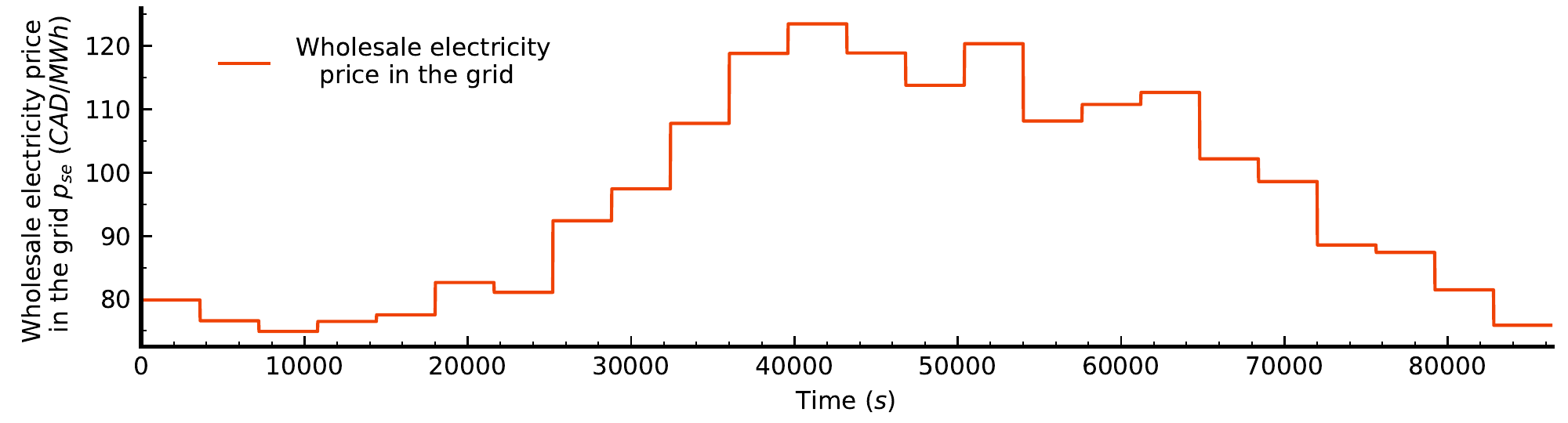}
		\caption{Wholesale electricity price in the grid.}
		\label{f11}
	\end{figure}
	
	For the following simulations, the time evolution of the external conditions in 24 hours is illustrated in Figure~\ref{f10} with the ambient temperature $t_a$ and solar radiation $S_{ra}$, the electric and cooling demands $P_d$ and $Q_o$ in the microgrid. The temporal resolutions of day-ahead and real-time curves are 1 hour and 1 minute, respectively. All the real-time curves obey Gaussian distribution and are randomly distributed within the about $\pm$6\% range of the respective day-ahead curves. The wholesale electricity price in the grid $p_{se}$ in a typical operating day is shown in Figure~\ref{f11}. The electricity price in microgrid $p_{mg}$ is set as 80 CAD/MWh. The natural gas price $p_f$ is 0.2 CAD/kg. The above prices refer to the relevant prices in Ontario, Canada, in the summer of 2021 \cite{ieso2021ieso, oeb2021oeb}. And assume that the compensation and fine by the grid, $p_{cm}$ and $p_{pn}$, associate with the wholesale electricity price $p_{se}$, and $p_{cm} = 1.5 p_{se}$, $p_{pn} = 1.5 p_{se}$.
	
	\subsection{Case 1: Offering 25\% available regulation capacity}
	
	\begin{figure}[!ht]
		\centering
		\includegraphics[width=0.85\hsize]{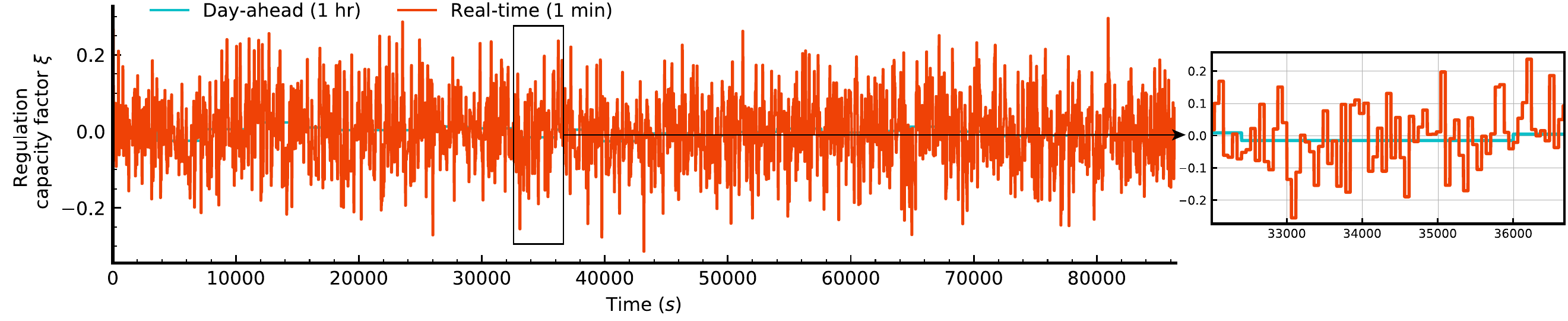}
		\caption{Time evolution of the regulation factor $\xi$ in Case 1: the day-ahead predicted factor used in the day-ahead scheduling; the actual real-time factor used in Problems 1-4.}
		\label{f12}
	\end{figure}
	
	To investigate Problems 1-4's performance, we apply Problems 1-4 to the IES, allowing it to provide about $\pm$25\% available regulation capacity for the grid. In this instance, the IES will send electricity to the grid according to its real-time instructions. If the grid needs more power than the planned baseline power $y_e^b$, the IES will increase the amount of power it sends, and if the grid needs less power than $y_e^b$, the IES will decrease the power it delivers. The adjustable range of the electricity supplied will be about 25\% above and below the baseline power level $y_e^b$, i.e., the regulation factor $\xi$ basically satisfying $\xi \in [-25\%, 25\%]$. The evolution of $\xi$ is depicted in Figure~\ref{f12}, in which the temporal resolutions of day-ahead and real-time curves are 1 hour and 1 minute. The real-time factor is distributed around the day-ahead prediction and obeys Gaussian distribution.
	
	\begin{figure}[!ht]
		\centering
		\includegraphics[width=0.86\hsize]{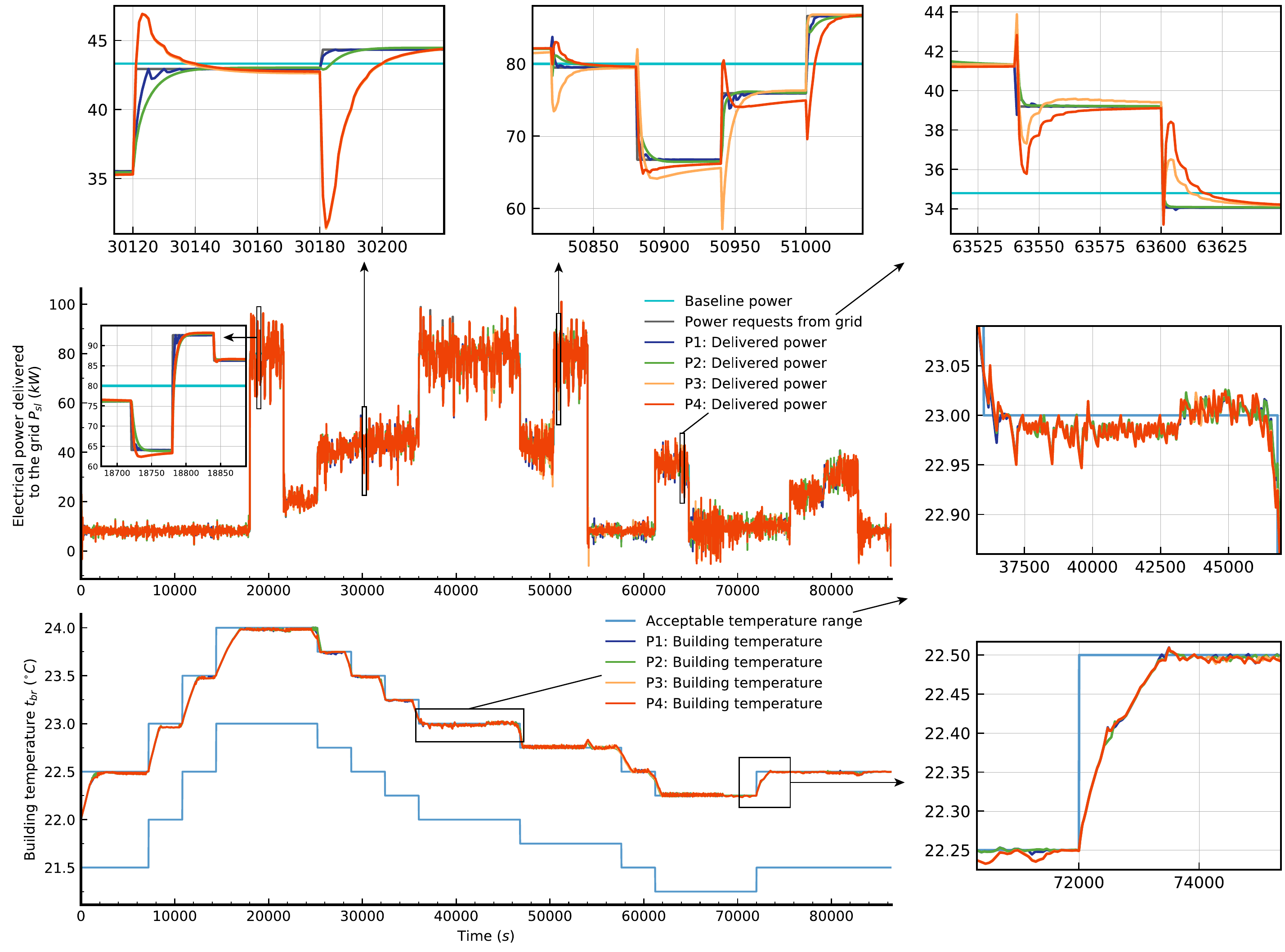}
		\caption{Case1: the power delivered to the grid (upper) and the building temperature (lower) under Problems 1-4.}
		\label{f13}
	\end{figure}
	
	\begin{figure}[!ht]
		\centering
		\includegraphics[width=0.82\hsize]{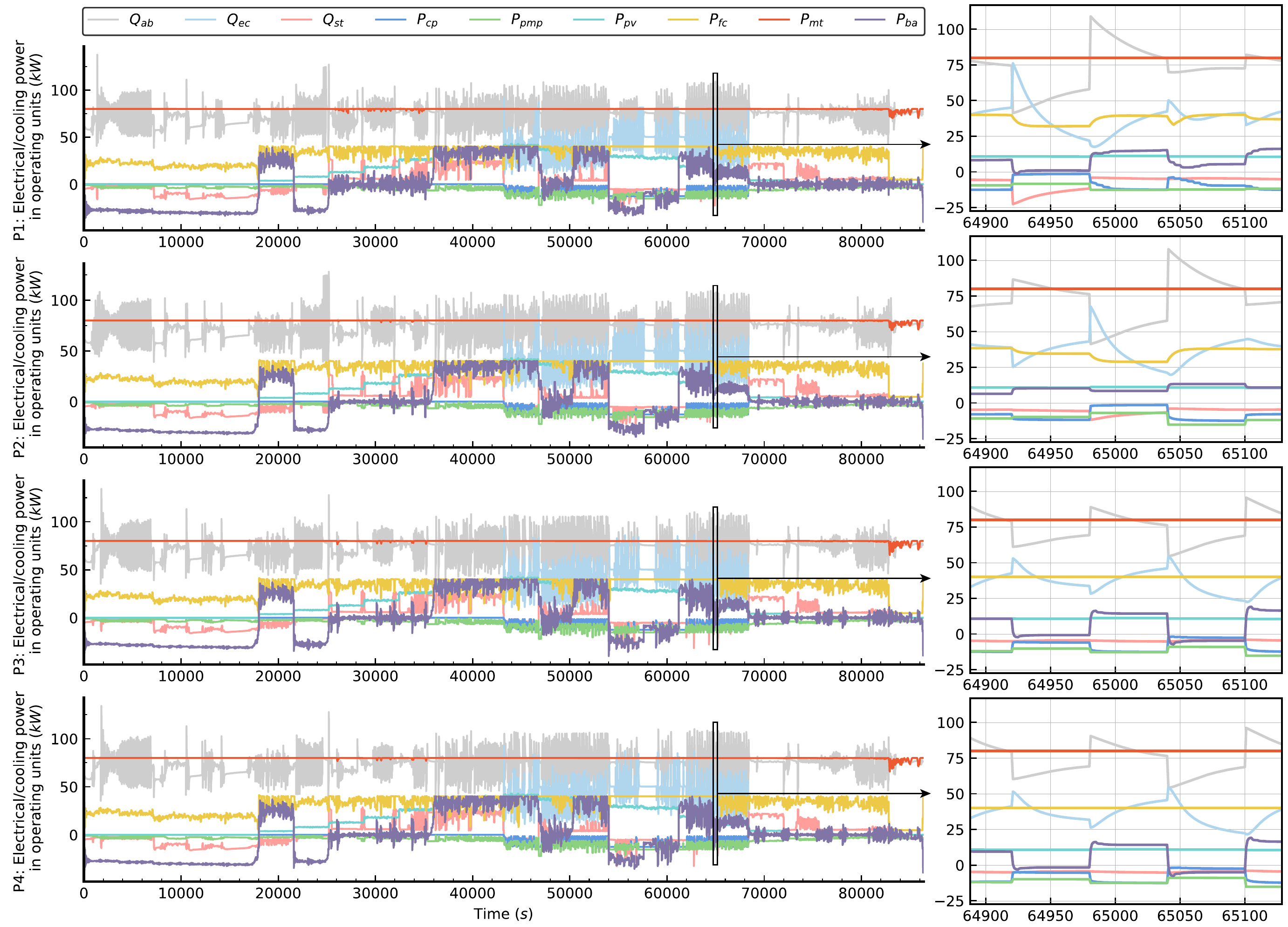}
		\caption{Case1: the generated electricity or cooling by each operating unit under Problems 1-4 (top-down).}
		\label{f14}
	\end{figure}
	
	\begin{figure}[!ht]
		\centering
		\includegraphics[width=0.5\hsize]{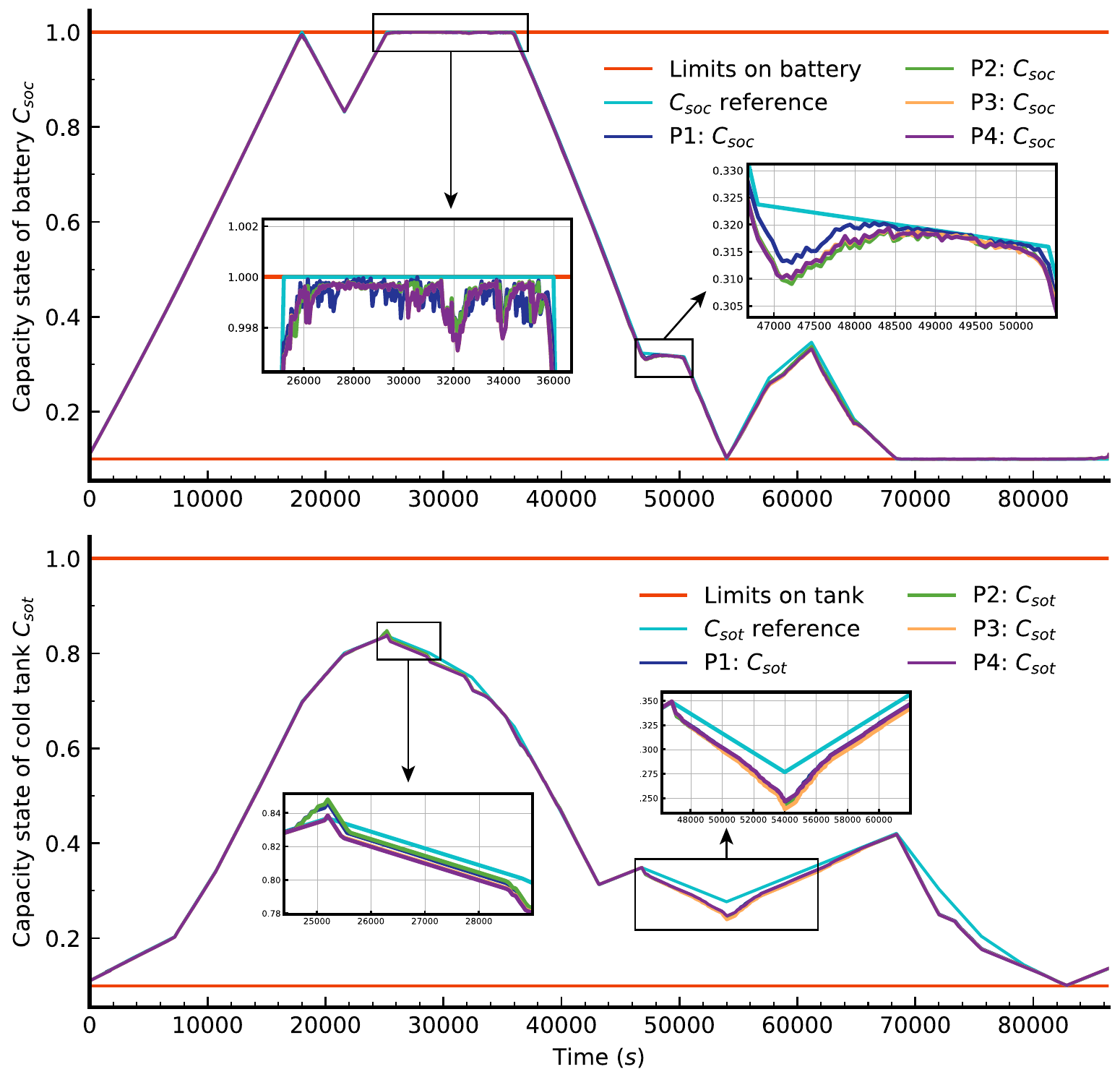}
		\caption{Case1: the capacity states of energy storage: battery bank (upper); cold storage (lower).}
		\label{f15}
	\end{figure}
	
	The simulation results in Case 1 are shown in Figures~\ref{f13}-\ref{f15}. As presented in Figure~\ref{f13}, the grid randomly requests the IES to regulate its sent power during operation. For example, the IES is required to increase its supplied power from about 65 kW to about 93 kW at 18780 s while the baseline power $y_e^b=80 \ {\rm kW}$. During such grid response, the proposed DEMPC, i.e., P1, exhibits superior performance in tracking the grid's real-time instructions for increasing/decreasing the sent power. P2, the supervisory MPC based on the developed subsystem configuration, is inferior to P1 but still outperforms P3 and P4 in terms of precise control of the supplied electricity. The supervisory MPC base on the empirical decomposition, P3 and P4, display comparatively poor performance, which cannot precisely track the changing power instructions, sometimes, cannot even reach the current prescribed power point before the next comes. Regarding the building temperature, Problems 1-4 basically satisfy the customers' demand for keeping the indoor temperature within the acceptable range. Furthermore, it can be seen that all these control schemes drive the building temperature to approach its upper bounds for energy and cost savings similar to \cite{halvgaard2012economic}.
	
	Figure~\ref{f14} portrays the dynamic conduct of each operating unit and their generated or consumed electricity or cooling power. We can observe that the units under Problems 1-4 have similar outlines of the power output since they use the same day-ahead scheduling. However, if we turn to the zoomed-in plots, we can find that the units under P1 are leveraged with more precise adjustments in the transient processes. In particular, the units exhibit complementary roles in dynamic behavior, i.e., synergy between the units, such as the fuel cell, battery, and compressor on the electricity side, and the absorption chiller, electric chiller, and cold storage on the cooling side. Moreover, from Figure~\ref{f15}, it can be seen that all these control strategies can tightly track the prescribed references for energy storage for long-term load shifting. It also proves that whichever controller is applied to the system, energy storage is neither overused nor underused in real-time control. The above observations demonstrate two key points: (a) the supervisory MPCs and proposed DEMPC have the capability to effectively manage the IES, but the proposed DEMPC and subsystem partition enhances the system's dynamic performance further; (b) the improvements are not attributed to a certain operating unit but a result of collaboration between all the units, which distinguishes the proposed method from what approaches are developed in existing work \cite{myovela2019feasibility}.
	
	To quantify the difference between Problems 1-4, we establish the following performance evaluation criteria based on the control objectives in Eq.\eqref{e5}:
	\begin{subequations} \label{e14}
		\begin{align}
			E_p &= \sum_{k=1}^{N_{sd}} J_1^{\frac{1}{2}}(k) \\
			E_t &= \sum_{k=1}^{N_{sd}} \Delta y_2(k) \\
			E_e &= \sum_{k=1}^{N_{sd}} - J_3(k) + \Delta C_{es} \\
			E_{glb} &= \beta_1 E_p + \beta_2 E_t - \beta_3 E_e 
		\end{align}
	\end{subequations}
	where $E_p$ accounts for the summation of the tracking deviation in the supplied electricity; $N_{sd}$ is the simulation duration; $E_t$ stands for the total tracking deviation in the building temperature, wherein $\Delta y_2(k)$ is the distance from the real indoor temperature to the acceptable range, i.e., if $y_2(k) \in [y_{sp,t}^{min}(k), y_{sp,t}^{max}(k)]$, $\Delta y_2(k) = 0$, otherwise, $\Delta y_2(k) = min\{\|y_2(k)-y_{sp,t}^{min}(k)\|, \|y_2(k)-y_{sp,t}^{max}(k)\|\}$; $E_e$ denotes the total profit, in which $\Delta C_{es}$ represents the cost owing to the unused energy stored in the storage units at the end of the operating day; $E_{glb}$ reflects the global performance of the system operation, where $\beta_1 = 0.05$, $\beta_2 = 3.5$ and $\beta_3 = 10$ are the normalization factors. For these indices, smaller $E_p$ and $E_t$ indicate better performance in the rapid response to the grid's requests and meeting the customers' cooling demand. Larger $E_e$ means higher economic gain. Smaller $E_{glb}$ represents superior overall operational performance.
	
	\begin{table}[!ht] 
		\centering
		\caption{Performance evaluation of Problems 1-4 in Case 1}
		\label{t5}
		\renewcommand{\arraystretch}{1.3}
		\tabcolsep 16pt
		\begin{tabular}{ccccc} \hline
			Case 1 & $E_p$ & $E_t$ & $E_e$ & $E_{glb}$ \\ \hline
			P1 & 10247 & 380 & 74.9 & 1093.7 (38.5\%) \\
			P2 & 28908.1 & 381.6 & 72.4 & 2056.9 (72.4\%) \\
			P3 & 41689.1 & 384 & 71.3 & 2715.7 (95.5\%) \\
			P4 & 41250.6 & 427.2 & 71.5 & 2842.2 (100\%) \\ \hline
		\end{tabular}
	\end{table}
	
	Table \ref{t5} displays the resulting performance indices of Problems 1-4 in Case 1. From the table, we can see that P1 is capable of significantly reducing the power deviation from the grid's real-time instructions while slightly improving the system's profitability. P2 exhibits suboptimal performance on system economics and precise control of the supplied power. Additionally, P1 is comparable in fulfilling customers' demand for cooling with P2 and P3. As a result, P1's overall performance surpasses P3 and P4 by about 60\%, and P2 outperforms P3 and P4 by about 25\%. These quantitative data illustrate the effectiveness and superiority of the developed subsystem decomposition method and distributed cooperation framework in the rapid response to the grid's requests and increasing profitability.
	
	\subsection{Case 2: Not participating in grid response}
	
\begin{figure}[!ht]
		\centering
		\includegraphics[width=0.86\hsize]{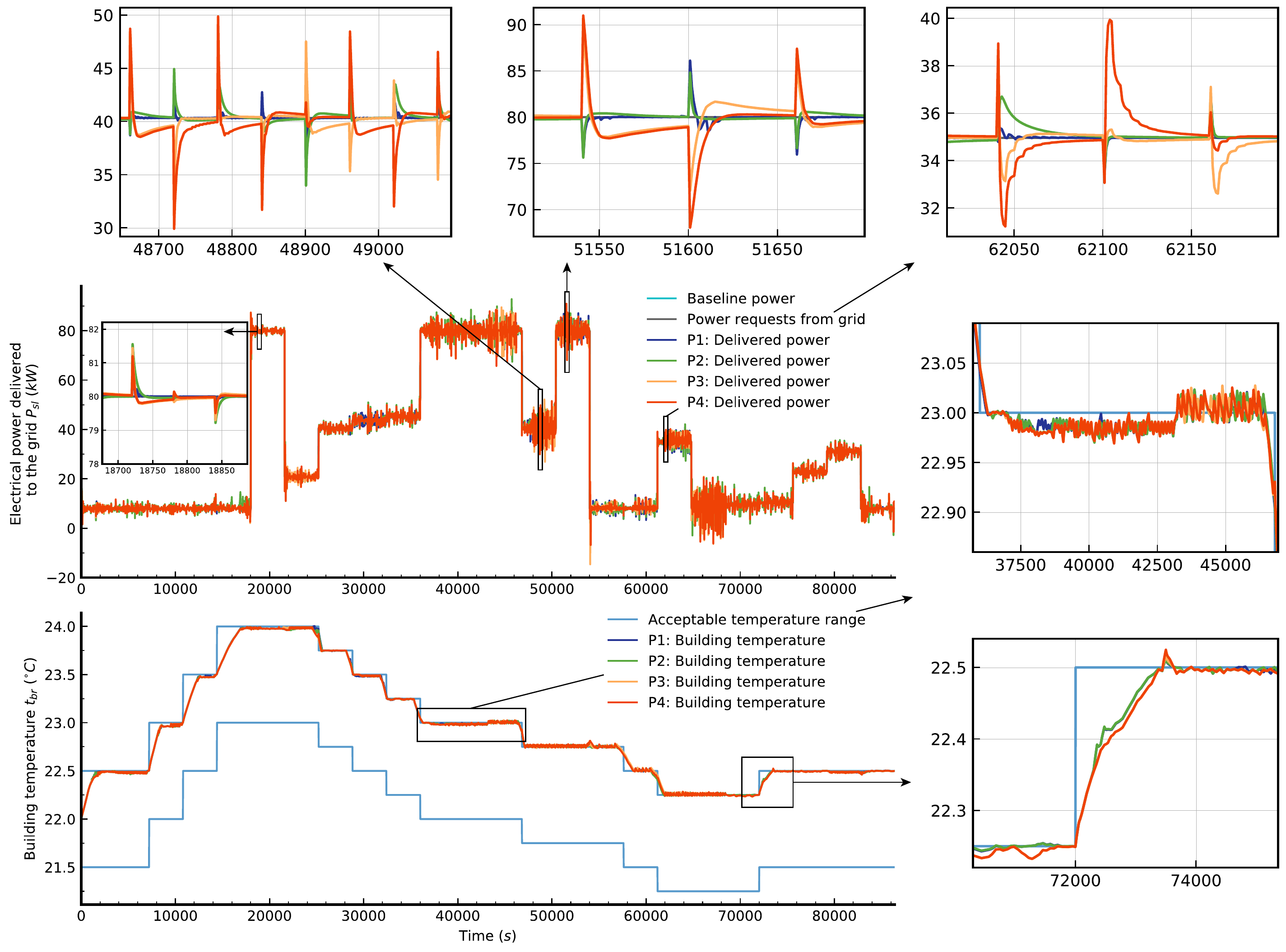}
		\caption{Case2: the power delivered to the grid (upper) and the building temperature (lower) under Problems 1-4.}
		\label{f16}
	\end{figure}

\begin{figure}[!ht]
		\centering
		\includegraphics[width=0.82\hsize]{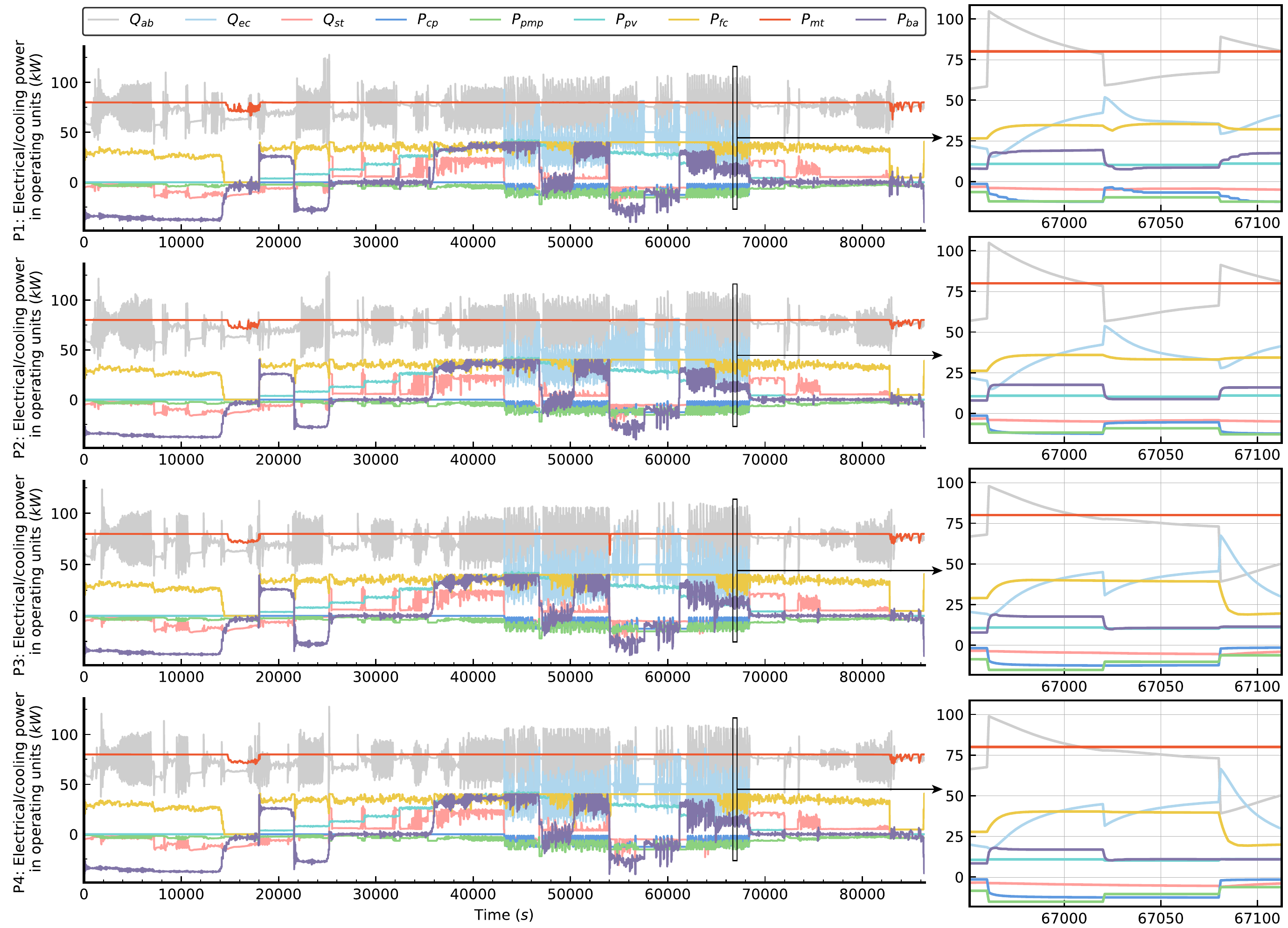}
		\caption{Case2: the generated electricity or cooling by each operating unit under Problems 1-4 (top-down).}
		\label{f17}
	\end{figure}

	In practice, the managers of IESs may also decide to send power to the grid according to the prior planned power baseline instead of the grid's real-time instructions. Then the regulation factor $\xi$ is always set as 0 during the operating day. In this case, the simulation results of Problems 1-4 are shown in Figures~\ref{f16} and \ref{f17}.
	
	From Figure~\ref{f16}, we can observe that P1 can basically keep the power sent to the grid with the baseline, while P2 is inferior to P1 but marginally preferable to P3 and P4. Even if the power baseline within an hour holds constant, the IES's exhibition of tracking the baseline is also degraded by P3 and P4 since they need to struggle against the frequent variation of the external conditions and local customers' demands without the system synergy. Meanwhile, the building temperature is mainly held within the acceptable range under all these control frameworks. As shown in Figure~\ref{f17}, the operating units under P1 are still operated in a dynamically complementary and collaborative manner in the transient processes like in Case 1. It is worth noting that since the IES is not required to respond to the grid's requests any longer, which diminishes system uncertainty, the electrical or cooling power fluctuations of the units are less than those in Case 1.
	
	\begin{table}[!ht] 
		\centering
		\caption{Performance evaluation of Problems 1-4 in Case 2}
		\label{t6}
		\renewcommand{\arraystretch}{1.3}
		\tabcolsep 16pt
		\begin{tabular}{ccccc} \hline
			Case 2 & $E_p$ & $E_t$ & $E_e$ & $E_{glb}$ \\ \hline
			P1 & 6786.1 & 346 & 65.9 & 891.4 (44.1\%) \\
			P2 & 22151.1 & 330.3 & 64.5 & 1618.8 (80.1\%) \\
			P3 & 27413.4 & 395.2 & 63.9 & 2114.5 (104.6\%) \\
			P4 & 27121.5 & 373.9 & 64.3 & 2021.4 (100\%) \\ \hline
		\end{tabular}
	\end{table}
	
	Table \ref{t6} lists the evaluation indices under Problems 1-4 in Case 2. Problem 1 exhibits outstanding performance in the electricity-relevant index, which indicates P1 accurately follows the power baseline while overcoming changing external conditions. The satisfaction of the indoor temperature requirement and the system profitability under P1 is better than P3 and P4 a bit. The performance of P2 also slightly surpasses P3 and P4 in all aspects. Consequently, the overall performance of the IES is boosted by about 55\% under P1 and about 20\% under P2. These results reveal that the developed subsystem partition and DEMPC remain effective and robust and enhance the system performance when the IES does not participate in the grid response. Furthermore, we also note that all the indices are less than those in Case 1. Electricity and cooling indices decline as the control systems do not need to consider the time-varying power instructions by the grid. However, losing compensation for the grid response, the system earnings also declines. It is important to note that the significance of the grid response lies not only in raising the IES economic gain but also in contributing to the grid's reliability and flexibility.
	
	\subsection{Case 3: Offering varying available regulation capacity}
	
\begin{figure}[!ht]
		\centering
		\includegraphics[width=0.86\hsize]{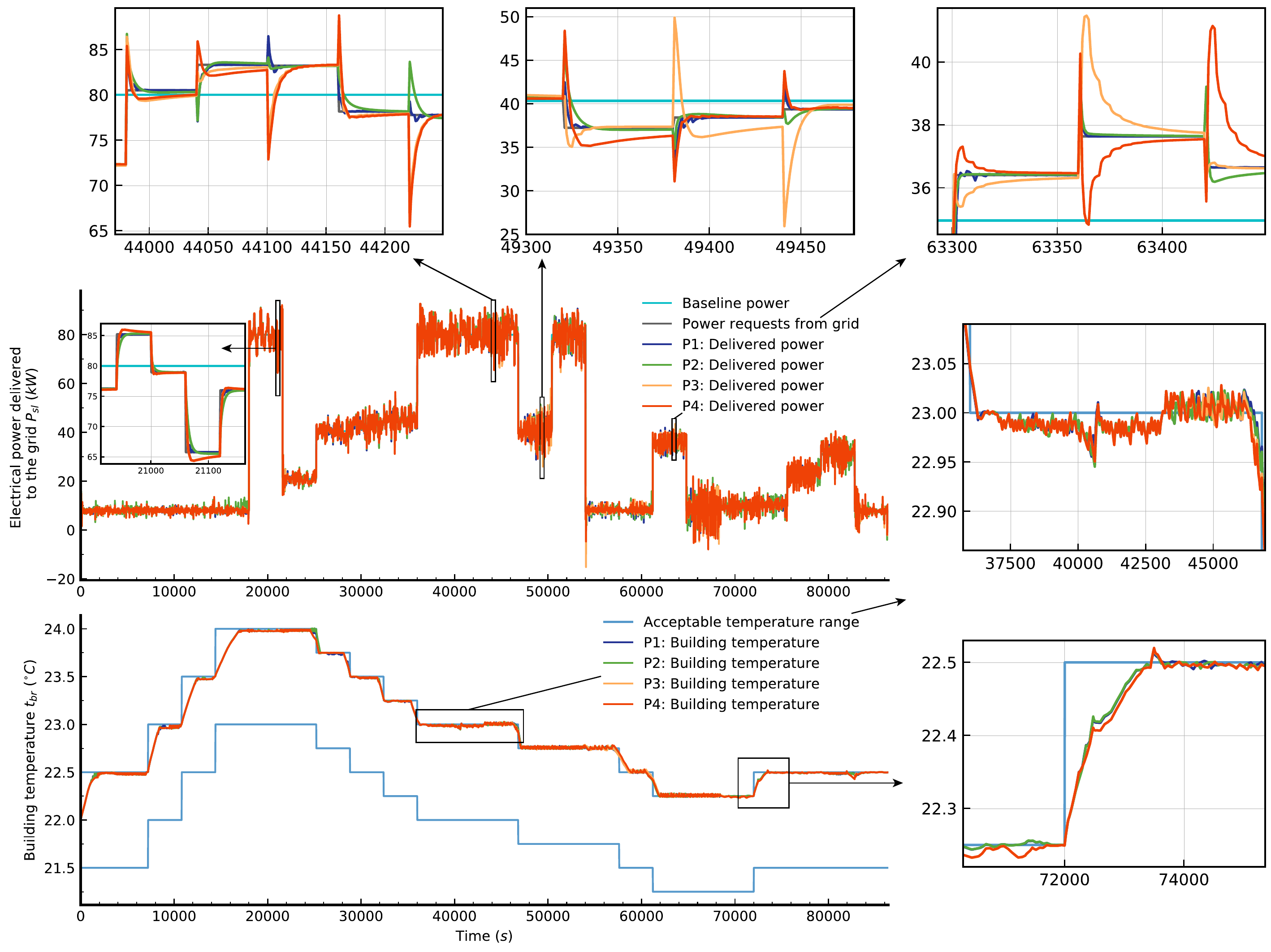}
		\caption{Case3 (available regulation capacity of 15\%): the power delivered to the grid (upper) and the building temperature (lower) under Problems 1-4.}
		\label{f18}
	\end{figure}

\begin{figure}[!ht]
		\centering
		\includegraphics[width=0.86\hsize]{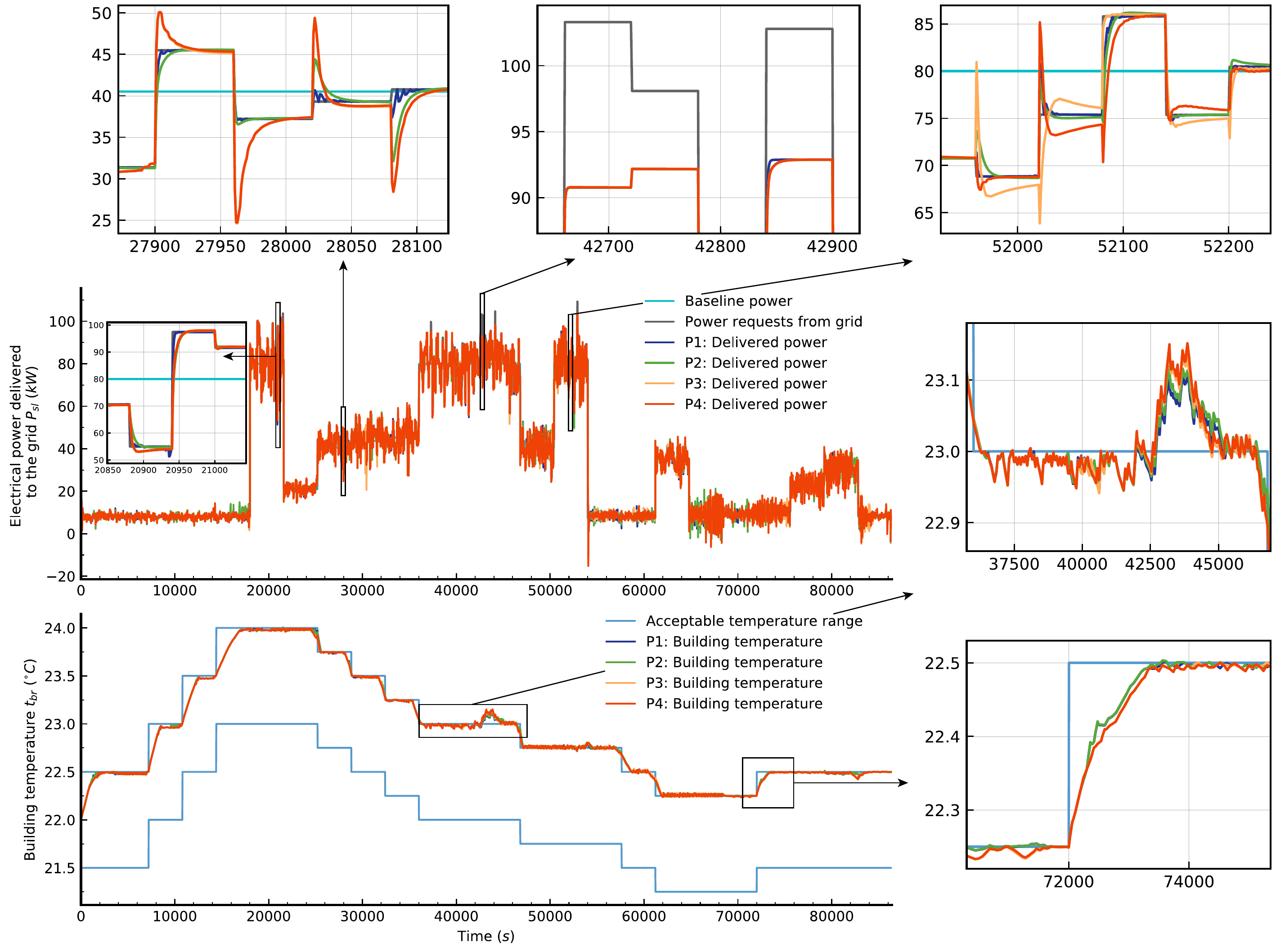}
		\caption{Case3 (available regulation capacity of 30\%): the power delivered to the grid (upper) and the building temperature (lower) under Problems 1-4.}
		\label{f19}
	\end{figure}

	To explore the system performance with different working conditions, we allow the IES to provide the grid with available regulation capacities from 0\% to 30\% to participate in the grid response. The instances with the available capacities of 25\% and 0\% have been discussed in Case 1 and 2. Figures~\ref{f18} and \ref{f19} depict the simulation results with the available capacities of 15\% and 30\%, respectively.
	
	The system behavior with the available capacities of 15\% and 30\% under Problems 1-4, as shown in Figures~\ref{f18} and \ref{f19}, are akin to its performance in Case 1. Due to the close collaboration between the operating units, P1 outperforms other controllers in immediate response to the grid's requests. P2 remains superior to P3 and P4 in tracking to the grid's real-time instructions owing to the proposed subsystem configuration. At the same time, Problems 1-4 perform alike in fulfilling the demand for maintaining the indoor temperature within the customer's desired range. In the case of the available regulation capacity of 30\%, we detect that the IES cannot meet the grid's demands for the supplied power occasionally, such as at 42700 s, since under given current external conditions, the grid's instructions for increasing/decreasing the supplied power may already exceed the limits of the system operation. In this particular case, the IES has to relax the requirement of the building temperature to regulate the generated electricity first as possible. Therefore, we can see that the indoor temperature has crossed the upper bounds at about 42700 s. Similar phenomena are also observed in other existing research on grid response, see \cite{wang2022machine}.
	
	\begin{figure}[!ht]
		\centering
		\includegraphics[width=0.72\hsize]{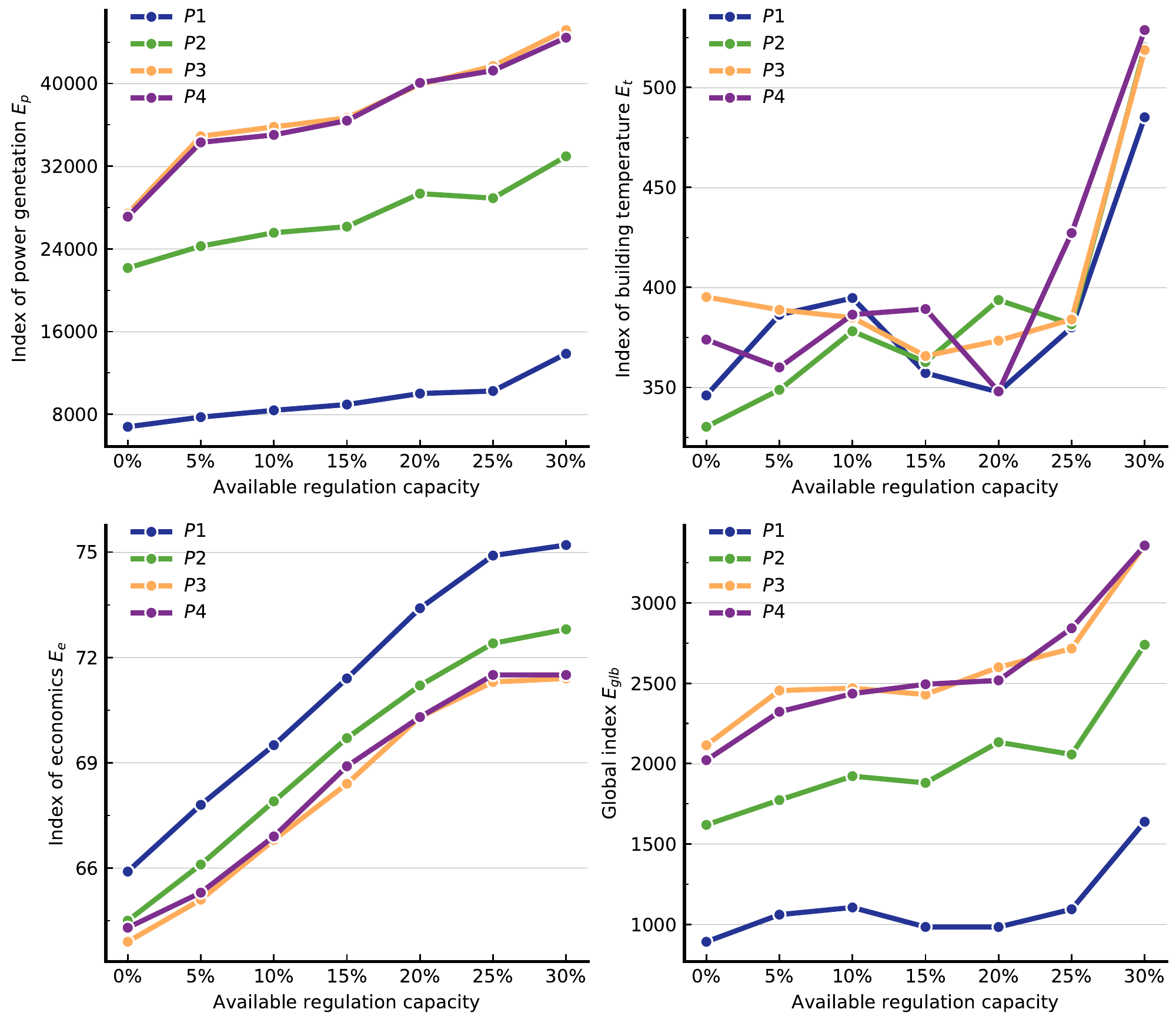}
		\caption{The evaluation indices under varying available regulation capacity: $E_p$ (up left), $E_t$ (up right), $E_e$ (down left), $E_{glb}$ (down right).}
		\label{f20}
	\end{figure}
	
	Figure~\ref{f20} describes changes of the evaluation indices over the available regulation capacity. From the figure, we can find that, first, as the growing available regulation capacity, the power deviation index $E_p$ under all the controllers displays an uptrend. This trend is expected since offering larger regulation capacity requires the controllers to have a greater capability to coordinate the entire system to precisely track the grid's dramatically changing instructions. In this regard, the proposed DEMPC substantially outperforms other controllers. The control framework based on the proposed subsystem configuration, P2, consistently surpasses P3 and P4, although they have the same control architecture as P2. Moreover, under P3 or P4, there is an apparent rise in $E_p$ when the available capacity is from 0\% to 5\%. It indicates that when the IES turns to participate in the grid response, the P3 or P4 based on the empirical decomposition is not robust regarding precise control of the generated power. Second, all the control strategies are similar in the index of meeting the cooling demand $E_t$. And a sharp rise in $E_t$ appears after the available capacity of 25\% due to sometimes $\xi$ being beyond the reach of the system. Regarding the system profitability $E_e$, as illustrated in the figure, the lager available regulation capacity the more earnings they can have because of the compensation for the grid response. The system revenue under P3 or P4 is slightly less than P2, while P2 is lower than P1. Additionally, the uptrend of $E_e$ slows down after the available capacity of 25\%, as the IES sometimes cannot respond to the grid's requests, hence, is fined more. Last, the IES's overall performance index $E_{glb}$ increases with the available capacity, which indicates the IES that provides a higher available capacity requires an outstanding control strategy to ensure its steady performance. As shown in the figure, the performance of P3 and P4 is inferior to P2. This degradation demonstrates the effectiveness of the proposed subsystem decomposition in achieving optimal modular management of the IES. The proposed DEMPC surpasses P2, which exhibits the superiority of the DEMPC in exploiting the potential of synergy between the operating units. Accordingly, the DEMPC has the superior capability to precisely control the supplied power at the grid's requests while raising the system's earnings and maintaining the building temperature.
	
	\begin{table}[!ht] 
		\centering
		\caption{Average iterations in the fast EMPCs}
		\label{t7}
		\renewcommand{\arraystretch}{1.3}
		\tabcolsep 8pt
		\begin{tabular}{cccccccc} \hline
			Available regulation capacity & 0\% & 5\% & 10\% & 15\% & 20\% & 25\% & 30\% \\ \hline
			Average iterations & 2.315 & 2.313 & 2.324 & 2.333 & 2.350 & 2.349 & 2.354 \\ \hline
		\end{tabular}
	\end{table}
	
	On the iteration times of the fast EMPCs during the simulations, the reached minimum and maximum iterations are 2 and 12, respectively. The mean iteration times of the fast EMPCs are also investigated in Table \ref{t7}. It can be seen that the mean iteration time is about 2.3 under multiple working conditions. This stable convergence of the DEMPC shows its applicability in practice.
	
	\section{Conclusions}
	
	IESs typically show the potential for greater reliability and flexibility of the grid. However, tight interconnections and interactions between various operating units in IESs are unfavorable for designing a proper real-time control scheme. To address the dynamic and structural complexity, we propose a systematic subsystem decomposition method based on a directed graph representation of IESs. By the proposed approach, the entire IES is decomposed vertically based on the dynamic time scale and horizontally based on the closeness of interconnections between the units. In addition, the qualitative analysis of decomposition reveals that, in the case of IESs, vertical decomposition should be carried out first to establish a consistent time scale within each subsystem and then horizontal decomposition. This order of decomposition is more conducive to designing distributed cooperation schemes for IESs. Based on this conclusion, we draw a control-oriented basic guideline for decomposing complex energy systems into optimal subsystems. Utilizing the decomposed subsystems, we develop a scalable cooperative DEMPC with global objectives for enhanced responsiveness while meeting the cooling and economic requirements. In the DEMPC, multiple local agents cooperate sequentially and iteratively in leveraging the units for the system-wide synergy.
	
	Extensive simulations demonstrate the applicability and effectiveness of the proposed subsystem decomposition and distributed cooperation framework. Whether or not the IES participates in the grid response, the control strategy based on the proposed subsystem configuration outperforms the same control architectures based on empirical decomposition. Due to collaboration between all the operating units, the developed DEMPC further significantly improves the system's dynamic performance, particularly on precise control of the generated power at the grid's requests. The investigations of the IES under changing working conditions exhibit that, compared with the empirical decomposition-based control, the proposed decomposition and cooperation scheme is more robust to cope with the IES providing multiple regulation capacities for the grid. Furthermore, we find that with the increasing available regulation capacity and the deepening grid response, the requirements for coordinated control systems become increasingly demanding. Effective control strategies will be essential for the precise, deep grid response. An overlarge regulation capacity beyond the reach of the system will lead to an apparent decline in overall performance of the IES, which also implies a deteriorated supplied power quality for the grid and a stagnant economy for the IES.
	
	\section{Acknowledgments}
	This work was supported by National Natural Science Foundation of China (Grant 51936003); National Key R\&D Program of China (Grant 2022YFB4100403); China Scholarship Council. The second author X. Yin would like to acknowledge the financial support from Ministry of Education, Singapore, under its Academic Research Fund Tier 1 (RS63/22), and Nanyang Technological University, Singapore (Start-Up Grant).


\end{document}